\documentclass[%
 reprint,superscriptaddress,
 amsmath,amssymb
 aps,
longbibliography
]{revtex4-1}

\usepackage[T1]{fontenc}
\usepackage[utf8]{inputenc}
\usepackage{times}
\usepackage{amsthm}
\usepackage{amssymb}
\usepackage{amsmath}
\usepackage{comment}

\usepackage{graphicx}
\usepackage{xcolor}

\definecolor{myrefcolor}{rgb}{0.067,0.5,0.5}

\usepackage[
    breaklinks,
    pdftex,
    colorlinks=true,
    linkcolor=myrefcolor,
    citecolor=myrefcolor,
    urlcolor=myrefcolor,
    linkcolor=myrefcolor,
    citecolor=myrefcolor,
    filecolor=myrefcolor,
    menucolor=myrefcolor,
    runcolor=myrefcolor
]{hyperref}

\usepackage{enumitem}         
\setlist[itemize]{left=0pt, labelsep=0.5em, topsep=0pt, partopsep=0pt, parsep=0pt, itemsep=0pt}

\usepackage{titlesec}
\titleformat{\section}[block]{\centering\normalsize\bfseries}{\thesection}{1em}{}
\titleformat{\subsection}[block]{\centering\normalsize\bfseries}{\thesubsection}{1em}{}
\titleformat{\subsubsection}[block]{\centering\normalsize\itshape}{\thesubsubsection}{1em}{}

\AtBeginDocument{%
    \newwrite\bibnotes
    \def\bibnotesext{Notes.bib}
    \immediate\openout\bibnotes=\jobname\bibnotesext
    \immediate\write\bibnotes{@CONTROL{REVTEX41Control}}
    \immediate\write\bibnotes{@CONTROL{%
    apsrev41Control,author="08",editor="1",pages="1",title="0",year="1"}}
     \if@filesw
     \immediate\write\@auxout{\string\citation{apsrev41Control}}%
    \fi
}%

\begin{document}

\title{Quantum computing and artificial intelligence: status and perspectives}
\date{\today}

\author{Giovanni Acampora} 
\affiliation{University of Naples Federico II, I-80126 Naples, Italy}

\author{Andris Ambainis} 
\affiliation{Center for Quantum Computer Science, Faculty of Computing, University of Latvia, LV-1586 Riga, Latvia}

\author{Natalia Ares}
\affiliation{Department of Engineering Science, University of Oxford, Parks Road, Oxford, OX1 3PJ, United Kingdom}

\author{Leonardo Banchi} 
\affiliation{Department of Physics and Astronomy, University of Florence, I-50019 Sesto Fiorentino (FI), Italy}
\affiliation{INFN Sezione di Firenze, I-50019, Sesto Fiorentino (FI), Italy}

\author{Pallavi Bhardwaj} 
\affiliation{SAP SE, Walldorf, Germany}

\author{Daniele Binosi} 
\affiliation{Quantum Community Network}
\affiliation{European Centre for Theoretical Studies in Nuclear Physics and Related Areas (ECT*, Fondazione Bruno Kessler), I-38123 Villazzano (TN), Italy}

\author{G. Andrew D. Briggs} 
\affiliation{Department of Materials, University of Oxford, Oxford OX1 3PH, UK}

\author{Tommaso Calarco} 
\affiliation{Quantum Community Network} 
\affiliation{Forschungs\-zentrum Jülich, D-52428 Jülich, Germany}

\author{Vedran Dunjko} 
\affiliation{Leiden Institute of Advanced Computer Science, N-2333 Leiden, Netherlands}

\author{Jens Eisert} 
\affiliation{Dahlem Center for Complex Quantum Systems, Freie Universit{\"a}t Berlin, D-14195 Berlin, Germany}

\author{Olivier Ezratty} 
\affiliation{EPITA Research Lab}
\affiliation{Quantum Energy Initiative}

\author{Paul Erker} 
\affiliation{Vienna Center for Quantum Science and Technology, Atominstitut, TU-Wien, A-1020 Vienna, Austria}
\affiliation{IQOQI Vienna, {\"O}AW , A-1090 Vienna, Austria}

\author{Federico Fedele}
\affiliation{Department of Engineering Science, University of Oxford, Parks Road, Oxford, OX1 3PJ, United Kingdom}

\author{Elies Gil-Fuster}
\affiliation{Dahlem Center for Complex Quantum Systems, Freie Universit{\"a}t Berlin, D-14195 Berlin, Germany}
\affiliation{Fraunhofer Heinrich Hertz Institute, 10587 Berlin, Germany}

\author{Martin G\"{a}rttner}
\affiliation{Institute of Condensed Matter Theory and Optics, Friedrich-Schiller-University Jena, D-07743 Jena, Germany}

\author{Mats Granath}
\affiliation{Department of Physics, University of Gothenburg, 41296 Gothenburg, Sweden}

\author{Markus Heyl}
\affiliation{Theoretical Physics III, Center for Electronic Correlations and Magnetism, Institute of Physics, University of Augsburg, D-86135 Augsburg, Germany}
\affiliation{Centre for Advanced Analytics and Predictive Sciences (CAAPS), University of Augsburg, Universitätsstr. 12a, 86159 Augsburg, Germany}

\author{Iordanis Kerenidis} 
\affiliation{Université Paris Diderot, F-75013 Paris, France}

\author{Matthias Klusch} 
\affiliation{German Research Center for Artificial Intelligence (DFKI), D-66123 Saarbr{\"u}cken, Germany}

\author{Anton Frisk Kockum}
\affiliation{Department of Microtechnology and Nanoscience, Chalmers University of Technology, 41296 Gothenburg, Sweden}

\author{Richard Kueng} 
\affiliation{Johannes Kepler University Linz, A-4040 Linz, Austria}

\author{Mario Krenn} 
\affiliation{Max Planck Institute for the Science of Light, Erlangen, Germany}

\author{Jörg~Lässig}
\affiliation{University of Applied Sciences Zittau/Görlitz, Brückenstraße 1, D-02826 Görlitz, Germany}
\affiliation{Fraunhofer IOSB-AST, Wilhelmsplatz 11, D-02826 Görlitz, Germany}

\author{Antonio Macaluso} 
\affiliation{German Research Center for Artificial Intelligence (DFKI), D-66123 Saarbr{\"u}cken, Germany}

\author{Sabrina Maniscalco} 
\affiliation{University of Turku, FI-20014 Turun yliopisto, Finland}

\author{Florian Marquardt} 
\affiliation{Max Planck Institute for the Science of Light and Friedrich-Alexander-Universit\"at Erlangen-N\"urnberg, D-91058 Erlangen, Germany}

\author{Kristel Michielsen} 
\affiliation{Forschungs\-zentrum Jülich, D-52428 Jülich, Germany}

\author{Gorka Mu\~noz-Gil} 
\affiliation{University of Innsbruck, A-6020 Innsbruck, Austria}

\author{Daniel Müssig}
\affiliation{Fraunhofer IOSB-AST, Wilhelmsplatz 11, D-02826 Görlitz, Germany}

\author{Hendrik Poulsen Nautrup} 
\affiliation{University of Innsbruck, A-6020 Innsbruck, Austria}

\author{Sophie A. Neubauer} 
\affiliation{Pyrosoma.ai}

\author{Evert van Nieuwenburg} 
\affiliation{Leiden Institute of Advanced Computer Science, N-2333 Leiden, Netherlands}
\affiliation{Leiden Institute of Physics, N-2333 Leiden, Netherlands}

\author{Roman Orus} 
\affiliation{Donostia International Physics Center, E-20018 Donostia, Spain}

\author{Jörg Schmiedmayer} 
\affiliation{Vienna Center for Quantum Science and Technology, Atominstitut, TU-Wien, A-1020 Vienna, Austria}

\author{Markus Schmitt} 
\affiliation{Forschungs\-zentrum Jülich, D-52428 Jülich, Germany}
\affiliation{University of Regensburg, D-93053 Regensburg, Germany}

\author{Philipp Slusallek} 
\affiliation{Saarland University, D-66123 Saarbr{\"u}cken, Germany}
\affiliation{German Research Center for Artificial Intelligence (DFKI), D-66123 Saarbr{\"u}cken, Germany}

\author{Filippo Vicentini}
\affiliation{CPHT and LIX, CNRS, Ecole Polytechnique, Institut Polytechnique de Paris, 91120 Palaiseau, France}
\affiliation{Collège de France, Université PSL, 11 place Marcelin Berthelot, 75005 Paris, France}
\affiliation{Inria Paris-Saclay, Bâtiment Alan Turing, 1, rue Honoré d’Estienne d’Orves – 91120 Palaiseau}

\author{Christof Weitenberg} 
\affiliation{Department of Physics, TU Dortmund University, 44227 Dortmund, Germany} 

\author{Frank K.~Wilhelm} 
\affiliation{Saarland University, D-66123 Saarbr{\"u}cken, Germany} 
\affiliation{Forschungs\-zentrum Jülich, D-52428 Jülich, Germany}


\maketitle

\tableofcontents

\section{Executive summary}

Two computing revolutions are currently in the making with \emph{artificial intelligence} (AI) and \emph{quantum computing}, with different levels of maturity and market footprints. While the European Union’s scientific position in both domains is significant, it is still squeezed between the US dominance and China’s increasing role, particularly with AI. However, as demonstrated by the emergence of DeepSeek, within the AI domain of large language models, innovation, scientific astuteness, and open source models can significantly alter the market balance.

Leadership in these domains comes from scientific excellence and the capability to create strong integrated software platforms, to build significant, scalable, and efficient computing infrastructure, and to spur the creation of as generic as possible use cases for end users, from enterprise and public services to consumer markets. A tight integration between academic research and industry R\&D is a fundamental enabler of scientific progress and market success.

Quantum computing is still a promise in the making, but its synergies with AI are already there and growing~\cite{RevModPhys.91.045002,DunjkoReview,QMLReview}. Many AI-driven techniques are already enabling significant progress in quantum computing research and industry developments, from optimizing qubit control and \emph{quantum error mitigation} (QEM)~\cite{Cai2023} strategies in the early stage \emph{noisy intermediate-scale quantum} (NISQ) regime~\cite{bharti_2021_noisy} to designing novel quantum algorithms~\cite{AshleyOverview}, highlighting the importance of mastering this dual discipline as an enabling technology.

Likewise and the other way around, early evidence of quantum advantage in faster computing times, better results, or the need for less training data for solving specific \emph{machine learning} (ML) problems illustrates how uniting AI and quantum resources could bring value, even though the required large-scale, \emph{fault-tolerant quantum computers} (FTQCs) likely needed to obtain clearer such advantage remain a longer-term objective.

Public investment in the convergence of AI and quantum computing could strengthen the EU competitiveness by building on the existing research excellence in both fields and accelerate the transition from laboratories to market applications. One strategic goal of employing quantum computing is to advance AI-based solutions for healthcare, finance, materials discovery, and security. It is a mid- to long-term target, but one that will determine industrial and scientific leadership in both domains. Likewise, the EU AI and quantum research landscape should encourage the development of both open source and commercial integrated software engineering platforms.

As international competition escalates, ensuring support for academic research and private-sector innovation at this intersection will not only secure the economic benefits associated with disruptive breakthroughs but also reinforce the EU’s position as a prominent global player in emerging deep-tech ecosystems. A carefully orchestrated funding strategy that spans fundamental research, talent cultivation, and technology-transfer incentives will ensure a robust pathway from visionary lab-scale projects to tangible, high-impact industry platforms and applications. Accordingly, this white paper lays out the scientific and applications landscape for consolidating AI and quantum computing disciplines by providing a research and use case agenda.

This white paper starts with describing how quantum computing could help develop innovative AI solutions, particularly in the ML spaces. This is a mid- to long-term effort. It is aligned with quantum computer hardware road maps.

We then cover the use cases of classical AI to empower research and developments of quantum technologies, focused on quantum computing and quantum sensing. This application domain of AI will mature. One important aspect is to ensure classical AI scales well as the requirements of quantum computing platforms will grow, as the domain progressively shifts from NISQ devices to FTQCs. One example is the critical role of ML-powered \emph{quantum error correction} (QEC) techniques~\cite{RevModPhys.87.307}. At last, it provides a longer-term research agenda to drive work in foundational questions related to how AI and quantum computing interact and benefit each other.

The white paper ends with a set of recommendations and challenges on the way to orchestrate the proposed theoretical work, align quantum AI developments with quantum hardware road maps, work on both classical and quantum resource estimates, particularly with the goal to mitigate and optimize energy consumption, orchestrate this upcoming hybrid software engineering discipline, and develop the European industry competitiveness while considering societal aspects.

\section{Introduction}

The convergence of \emph{artificial intelligence} (AI) and \emph{quantum computing} is a rapidly evolving field, named \emph{quantum AI} (QAI), 
with the potential to impact numerous aspects of technology and science. This white paper explores the synergistic relationship between these two disciplines in QAI, outlining how quantum computing can enhance classical AI capabilities and how AI can be used to advance quantum computing. It details different methods, applications, and objectives, with a focus on both immediate and long-term goals. At this stage, it can be viewed as an early proposal for a strategic and industry research roadmap.

This document has been written by experts in AI and \emph{quantum technologies} (QT) in the spirit of the \emph{European Strategic Research and Innovation Agenda} (SRIA), that is trying to avoid overselling. Nevertheless, some parts of the text are speculative, due to the great novelty of the subject. 

Quantum AI covers all subfields of AI, such as \emph{quantum machine learning} (QML), \emph{quantum reasoning} (QR), \emph{quantum automated planning and scheduling} (QPS), \emph{quantum natural language processing} (QNLP), \emph{quantum computer vision} (QCV), and \emph{quantum agents and multi-agent systems} (QMAS) ~\cite{klusch2024quantum}. For example, in quantum-assisted ML, quantum processors may pre-process classical data that then feed classical ML methods. This approach may lead to improvements in total processing speed, accuracy, and reduce the amount of training data required. Quantum computing is also explored to accelerate the training phase of classical ML models, using both near-term \emph{noisy intermediate-scale quantum} (NISQ) devices 
\cite{bharti_2021_noisy} with variational algorithms~\cite{Variational}, and future \emph{fault-tolerant quantum computers} (FTQCs). Another approach is learning with quantum models, where quantum computing takes the helm for both the training and inference phases, potentially uncovering data patterns that are intractable for classical systems. 

Furthermore, quantum computing could significantly improve \emph{reinforcement learning} (RL)~\cite{meyer2022survey,sutton1998reinforcement}, addressing computational bottlenecks and lengthy training times. \emph{Quantum reinforcement learning} (QRL) methods in QML can use parameterized quantum circuits to optimize decision-making in complex environments, particularly in industrial applications. In unsupervised learning, quantum algorithms are developed to handle tasks like automatic clustering and dimensionality reduction, with the potential to provide exponential or polynomial speed-ups compared to the best classical methods available. This can for example be useful to develop innovative cyberthreat detection solutions.

These combined technologies, AI and quantum computing, have broad potential applications. In healthcare and life sciences, quantum simulations may be used to generate training data for AI models, which could accelerate drug discovery by exploring new chemical spaces. Medical image analysis used for X-rays, MRI scans, and biological sample imaging could be enhanced, reducing reliance on large and labeled data sets. In time-series analysis, quantum methods may provide more efficient modeling and faster anomaly detection. Quantum computing is also being explored for extracting insights from complex quantum systems.

AI is also being used to advance quantum computing through the design of novel quantum algorithms and protocols, the optimization of quantum circuits, with quantum error mitigation, and the implementation of quantum error correction, particularly for the costly task of error-syndrome detection. AI could help in discovering and optimizing quantum experiments, simulating quantum systems, and analyzing quantum data. We can even envision creating fully quantum AI models, where all data, training algorithms, and inference systems are quantum in nature.

The field faces several challenges, including current hardware limitations such as qubit numbers, fidelities, and scalability. There are difficulties in loading and processing classical data into quantum states at various levels. Training quantum models also presents challenges such as barren plateaus and the lack of efficient quantum equivalents to classical back-propagation used in the training of neural networks. Additionally, ensuring trust, robustness, interpretability, and explainability of AI models as well as avoiding various data biases, are critical for the reliable application of these technologies in practical situations. Standardized interfaces should be developed to share data and translate quantum problems into a common ML language.

\begin{figure}[t]
\centering
\includegraphics[width=8.5cm]{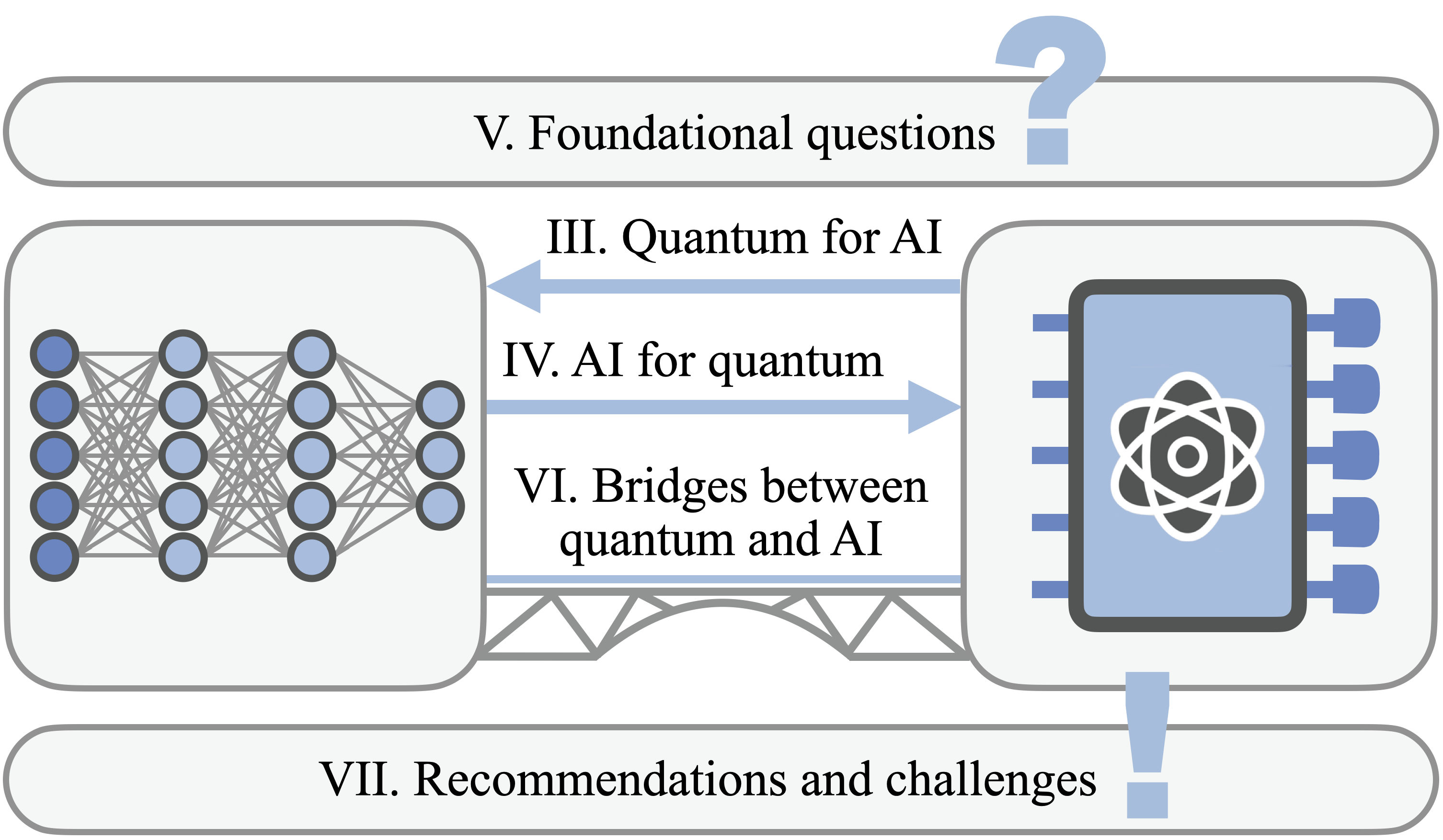}
\caption{An overview of the structure of this white paper.}
\label{Structure}
\end{figure}

This white paper outlines both short-term research goals (3 to 5 years), mid-term research goals (5 to 10 years), and long-term research goals (beyond 10 years) goals related to these various challenges. For example, in the short term, the focus is on demonstrating quantum utility for chemistry and error mitigation, identifying features that are easier to extract using quantum machines, and using AI to rediscover known quantum algorithms (see Fig.~\ref{Structure}).

Medium-term objectives include establishing hybrid frameworks for molecular simulations, new materials development, scaling hybrid classical-quantum models, and improving quantum error-correction techniques. Long-term goals involve validating frameworks for drug re-targeting tasks, broadening the scope of quantum-enhanced ML, developing fully quantum AI models, and using AI to co-design quantum algorithms and quantum hardware.

Finally, this white paper also emphasizes the importance of foundational questions about learning in a quantum world. These include understanding the connections between physics and QML, defining learning with quantum data, and exploring the limitations of quantum autonomous agents. The need for interdisciplinary collaboration, open-source software, and standardized data sets is also highlighted, along with the need to train the next generation of experts in both quantum information science and ML.

In conclusion, the combination of quantum computing and AI has the potential to drive significant advances across many sectors. Strategic research, addressing key challenges, and fostering collaboration will be of paramount importance for realizing the full potential of these technologies to the benefit of society and European competitiveness in both the AI and quantum computing fields.

\section{Quantum for AI}

The rapid advancements in AI across scientific and industrial domains have underscored the need to overcome the computational limitations of classical methods and explore alternative paradigms for scalable and efficient AI solutions. Recent progress in quantum computing suggests the potential for quantum-enhanced AI approaches to outperform purely classical techniques, 
particularly in addressing computationally intensive tasks~\cite{klusch2024quantum,ciliberto2018quantum}. A key direction in this integration is the development of hybrid quantum-classical architectures, where quantum processors serve as pre-processing units for classical AI inference tasks~\cite{Liu2024, Chen2025}. In the near to mid-term, this integration is expected to be feasible with NISQ devices comprising 100 to 200 physical qubits, while in the long term, early-stage FTQCs with over 50 logical qubits could enable more complex algorithms.

Quantum-enhanced AI primarily revolves around accelerating specific subroutines, such as for optimization, sampling, and high-dimensional data processing, which are computationally expensive for classical methods. One proposed approach involves combining quantum processors with \emph{high-performance computing} (HPC) resources into hybrid systems that leverage quantum algorithms for specific computational bottlenecks while retaining classical AI’s robustness and scalability. Another promising avenue in QML is the use of quantum-generated data to enhance ML models, potentially improving processing speed, computational complexity, modeling accuracy, and the amount of data required for training. While these developments highlight the transformative potential of quantum-assisted ML, realizing practical impact requires sustained interdisciplinary efforts from both the classical ML and quantum computing communities. 
That also applies accordingly to the other Quantum AI subfields such as quantum planning and scheduling, quantum reasoning, quantum natural language processing, quantum computer vision, and quantum multi-agent systems. 

\begin{figure}[h]
\begin{center}
\includegraphics[width=.9\columnwidth]{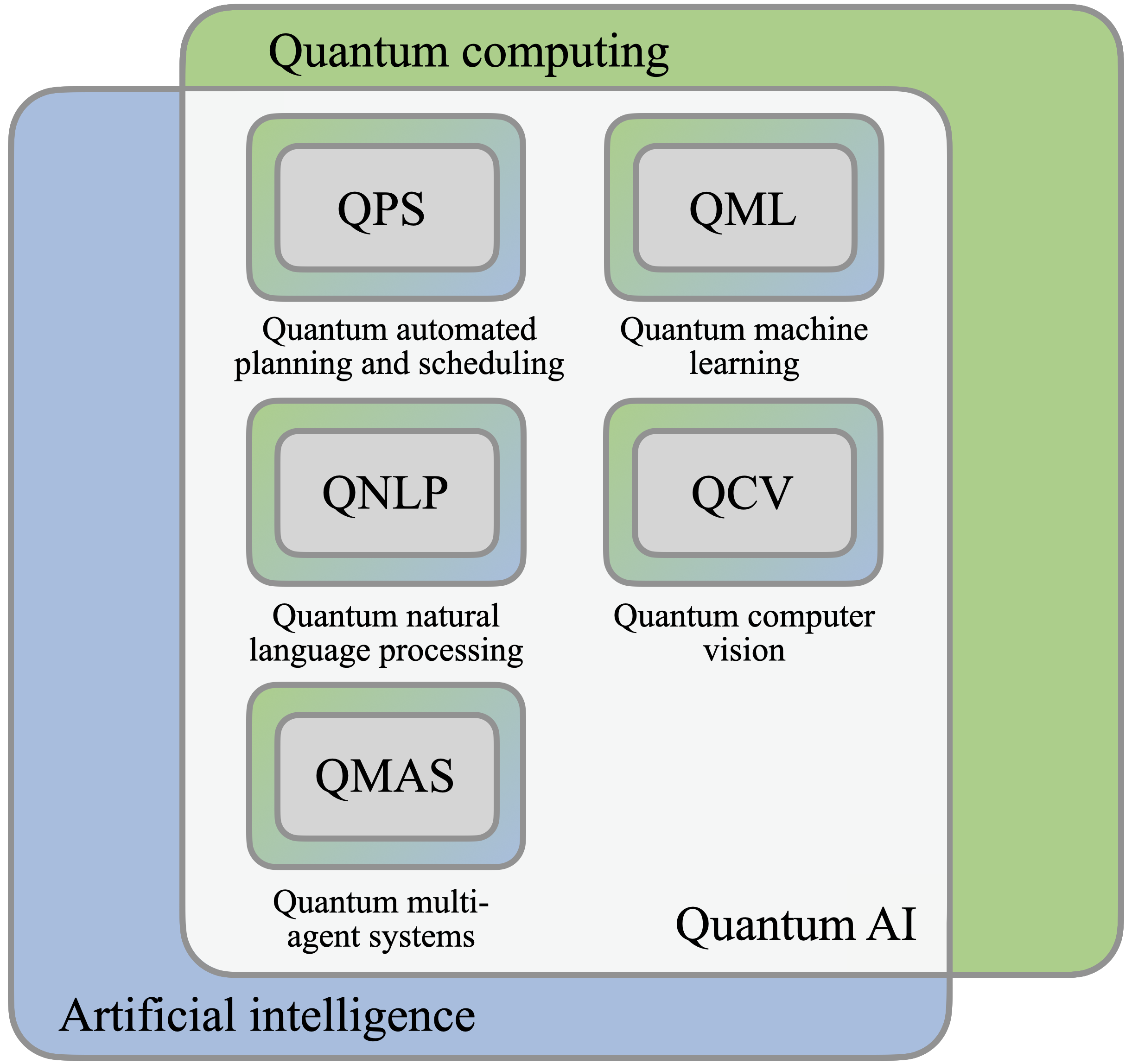}
\caption{\emph{Quantum AI} (QAI) as the intersection of quantum computing and AI with subfields in relation to AI 
each covering both directions. Image adapted from 
Ref.~\cite{klusch2024quantum}.}
\label{fig:quantum4ai}     
\end{center}
\end{figure}

This section briefly delineates the current state of quantum computing methods and their integration with classical AI workflows, with an emphasis on ML techniques. It further identifies research directions for other quantum AI areas — such as optimization, multi-agent systems, and reasoning — that currently remain underexplored despite the significant promise of quantum technologies.
This conceptual framework is illustrated in Fig.~\ref{fig:quantum4ai}.

The goals are the following:
\begin{itemize}
    \item Demonstrate quantum utility from using quantum processors as a pre-processing stage for classical AI inference tasks or full end-to-end QML solutions.
    \item Demonstrate this at scales achievable in the near- to mid-term with 100 to 200 physical qubits using quantum error mitigation and variational circuits (NISQ) or in the long-term early-stage FTQCs supporting between 50 and 100 logical qubits, which require a much larger number of physical qubits and allow much deeper quantum circuits.
\end{itemize}
The proposed scenarios are the following:
\begin{itemize}
    \item Quantum processors and HPC combined into hybrid systems, including hybrid algorithms.
    \item AI using data produced from a quantum processor, achieving overall improvement in
    \begin{itemize}
        \item processing speed and computational complexity,
        \item modeling capability and response accuracy,
       \item number of samples needed for training.
    \end{itemize}
    \item Classical and quantum resource estimation to identify expected usefulness and timeline.
\end{itemize}

\subsection{Quantum-assisted machine learning}

\subsubsection{Supervised learning}

\emph{Quantum supervised learning} (QSL)~\cite{quantumSL2024} refers to the application of quantum algorithms to solve supervised learning tasks. Supervised learning is a fundamental area of ML where a model is trained on labeled data to learn the relationship between inputs (features) and corresponding outputs (labels). The goal is to generalize this learned relationship to predict the outputs for unseen inputs. Common examples of supervised learning tasks include classification, where the model assigns an input to a specific category, and regression, where the model predicts a continuous quantitative value.

State-of-the-art classical methods, such as neural networks, have demonstrated remarkable performance in many tasks, particularly when large data sets and specialized hardware are available. However, these methods face several challenges, including the need for extensive sets of labeled data, long training times for complex models, limited explainability of predictions, and the high computational cost of modeling high-dimensional, non-linear relationships. These limitations often hinder the scalability and practicality of classical approaches in real-world scenarios.
Quantum supervised learning aims to address these challenges, tackling problems that are difficult or impossible for classical methods. In general, using quantum algorithms for supervised learning problems has different scopes: to simplify the training procedure in terms of training time and the number of optimization cycles required; to improve performance in terms of accuracy --- how well the models learn general patterns from the data; to enhance efficiency by reducing the amount of data needed for training, striving to achieve comparable or superior results with less information. This field encompasses a range of approaches that consider integrating quantum technologies into ML workflows, categorized by the extent of quantum involvement and the type of quantum hardware used. The distinction between FTQC and NISQ technologies, where fault tolerance is not assumed to be available, shapes the methodologies and practical applications of QSL, influencing both their theoretical potential and real-world feasibility~\cite{quantumSL2024,biamonte2017quantum}.

\medskip
\noindent\emph{Efficient quantum training of classical models}
\medskip

One key methodological approach of quantum computing in supervised learning is its use to enhance the training phase of classical models. Fault-tolerant QML~\cite{schuld2021fault} aims to achieve a theoretical speed-up in optimizing well-defined classical algorithms, such as support vector machines~\cite{rebentrost2014quantum}, splines~\cite{macaluso2020quantum}, and linear regression~\cite{lloyd2013quantum} by leveraging quantum algorithms to solve the underlying parametric optimization problems more efficiently.

These methods assume a specific functional form for the target function of interest, often expressed as a linear relationship with parameters to be estimated, such as in least-squares regression, thus enabling the use of quantum techniques like variants of the \emph{Harrow--Hassidim--Lloyd} (HHL) algorithm~\cite{PhysRevLett.103.150502} to accelerate computationally expensive linear algebra operations. These methods depend on the availability of error-corrected qubits and long quantum circuits, which enable efficient solutions to convex optimization problems with polynomial complexity. Strong guarantees for the performance of quantum training of classical models have in particular been found for models which are both sufficiently dissipative and sparse, as they arise in the context of quantum-training-pruned classical networks~\cite{Liu}.

In parallel, NISQ-based approaches, such as shadow models~\cite{jerbi2024shadows}, aim to leverage the current generation of quantum hardware to train classical surrogates of quantum neural networks that mimic quantum models under the assumption that training of these can be performed more efficiently~\cite{Variational}.

Additionally, in the neural-network-based deep learning realm, classically inspired quantum techniques, like \emph{quantum convolutional neural networks} (QCNNs)~\cite{QuantumConvolutional} or quantum perceptrons~\cite{kapoor2016quantum}, adapt established classical architectures to leverage computational advantages during training. 
All these approaches confine the use of quantum resources to the training phase, enabling classical inference for scalability and broader end-user applicability. However, in these cases, ad hoc techniques must be employed to ensure that the quantum component is used only during training, while keeping the algorithm entirely classical during testing. One such approach is knowledge distillation, where a quantum-trained model guides the training of a purely classical model. To transfer knowledge from a \emph{quantum neural network} (QNN)~\cite{QNN,QNN2} to a classical neural network, first, a variational-quantum-circuit-based QNN is trained on a classification or regression task, obtaining softmax-like output probabilities. Next, a classical neural network is trained using the soft labels predicted by the QNN instead of the ground truth. The goal is to generate a classical model that mimics the QNN’s decision boundary, effectively capturing its learned representations. This approach is particularly useful when classical models struggle with optimization, generalization, or data efficiency.

For instance, in image classification tasks, deep classical networks often require large labeled data sets and extensive hyperparameter tuning to achieve efficient convergence. Similarly, in time-series forecasting, classical models face challenges with long-range dependencies and high-dimensional correlations, necessitating extensive feature engineering and computationally expensive training. In molecular property prediction, classical models may struggle to capture quantum-mechanical properties using traditional feature representations, leading to suboptimal performance.

\medskip
\noindent\emph{Learning with quantum models}
\medskip

Beyond training enhancements, learning with quantum models incorporates quantum computing into both the training and inference phases, aiming to uncover patterns in data that are intractable for classical systems. This involves leveraging complex quantum kernels~\cite{schuld2019quantum} and parameterized quantum circuits~\cite{benedetti2019parameterized} to represent data in ways classical models cannot. The variational paradigm, which relies on hybrid quantum-classical techniques in which a smaller variational quantum circuit is controlled by a classical algorithm, is particularly prominent in NISQ-based implementations, offering opportunities to explore novel hypothesis functions~\cite{Variational}.

Preliminary results indicate the potential of quantum models to reduce the parameter space, require less training data, and enable more efficient training procedures~\cite{abbas2021power,huang2022quantum,caro2022generalization,Generalization}. These results currently hold only for specific scenarios involving \emph{quantum data}~\cite{PowerOfData}. Investigating these advantages in classical data-intensive domains, such as genomics (e.g., DNA-sequence analysis), financial modeling (e.g., high-frequency trading risk assessment), climate modeling (e.g., large-scale weather prediction), and industrial Internet of Things (e.g., real-time sensor network optimization), could open the possibility for quantum models to replace classical ones in both training and inference.

However, significant challenges persist, including barren plateaus in optimization~\cite{mcclean2018barren}, traps~\cite{anschuetz2022quantum}, and the lack of efficient quantum equivalents to the classical backpropagation training technique~\cite{abbas2023quantum}. In some instances, the quantum models can even be fully dequantized~\cite{Tang2022}, referring to the situation in which an efficient classical algorithm for the same task can be found, or classical surrogates formulated,
so classical models which can be efficiently obtained from a trained quantum learning model and reproduces its input-output relations \cite{PhysRevLett.131.100803}. 
At the same time, one has to ensure that the quantum circuit is sufficiently expressive. Despite these obstacles, the integration of quantum computing across the supervised learning pipeline holds huge potential by addressing computational bottlenecks and revealing new insights into complex data structures.

Diffusion (probability) models can be seen as a form of supervised learning. These models, specifically those used in generative tasks (like image generation), are based on a process where data is gradually corrupted with noise in a forward diffusion process, and then the model learns to reverse this noise process to recover the original data in the reverse diffusion process. Again, variants of the HHL algorithm can provide quantum algorithms for diffusion models for which there is evidence for a quantum advantage~\cite{QuantumDiffusionModels}.

\subsubsection{Reinforcement learning}

\emph{Reinforcement learning} (RL)~\cite{meyer2022survey,sutton1998reinforcement} is a branch of ML where an agent learns to make decisions by interacting with an environment to maximize cumulative rewards, outside the classification of supervised and unsupervised learning~\cite{RL}. It is used in many fields, e.g., robotics and large language models. Unlike supervised learning, which relies on labeled data, RL focuses on trial-and-error learning, using feedback from actions to improve future performance. A key challenge in RL lies in its computational complexity, as many RL problems are formulated using \emph{partially observable Markov decision processes} (POMDPs), which are computationally demanding. This complexity makes finding optimal solutions computationally intractable for large-scale problems, further highlighting the need for efficient algorithms and approximations~\cite{ding2020challenges,dulac2021challenges}. Practically, RL relies on extensive data sets and lengthy training times, particularly in state-of-the-art RL models which use deep neural networks. These hurdles underscore the potential of quantum computing to address the computational bottlenecks inherent in RL.

\medskip
\noindent\emph{Efficient quantum training of classical models}
\medskip

One promising avenue involves using quantum computing to enhance the training of classical reinforcement learning models by employing hybrid quantum-classical architectures. Many RL applications, such as robotic manipulation and navigation tasks, rely on \textit{actor-critic frameworks}~\cite{konda1999actor,grondman2012survey}, where the critic evaluates actions to stabilize policy updates for the actor. The critic’s role is vital in these systems, as accurate value estimation improves learning stability and convergence rates.

However, achieving this in complex environments demands high computational resources and extended training times. Hybrid architectures~\cite{kolle2024quantum,sinha2025nav} address this by integrating quantum neural networks, which act as the critic, with classical networks for the actor. Quantum neural networks used as the critic are expected to better capture high-dimensional patterns, better trainability and improved stability, as well as better generalization during the critic’s evaluation. Once training is complete, the quantum critic is removed, leaving a fully classical deployment system, without the need of a quantum computer during test. This design maximizes the practical utility of RL models while leveraging potential quantum advantages where they are most impactful, during training.

In actor-critic architectures (classical), the critic is only used at time of training, but the \emph{neural network} (NN) which outputs the action is the actor only. So, in this case there is no ``transfer'', because the quantum critic plays a role only in the training phase in both cases, classical and quantum. 

\medskip
\noindent\emph{Learning with quantum models}
\medskip

Another significant approach concerns quantum systems that replace the RL agent entirely~\cite{meyer2022survey}, requiring quantum computation during both training and inferences. In this context, \emph{parameterized quantum circuits} (PQCs)~\cite{Variational} act as quantum agents, processing information and optimizing decisions in environments where classical agents struggle. These quantum agents are anticipated to be particularly advantageous in scenarios involving optimization in high-dimensional spaces such as those encountered in protein folding optimization, where the state space is massive and training a traditional RL agent is extremely challenging. They may also address problems with inherent quantum properties, such as quantum synthesis or compilation---optimizing the execution of a quantum algorithm on specific quantum hardware.

While this approach holds transformative potential, it also faces significant challenges. These include the current limitations of quantum hardware, issues with scalability, and the practicality of deploying quantum computers in contexts where the agent must be dynamic and actively interact with the environment. Nonetheless, quantum reinforcement learning agents represent a frontier in combining quantum computational power with RL's adaptive frameworks to solve problems beyond the reach of classical methods.

\subsubsection{Unsupervised learning}

Quantum unsupervised learning seeks to leverage the principles of quantum computing to tackle the challenges of unsupervised learning tasks, which encompass clustering, dimensionality reduction, and generative modeling. One of the key distinctions within this field lies in the type of algorithms being developed: some aim to accelerate existing classical routines, while others introduce entirely new quantum-native methods. For instance, the $q$-means algorithm~\cite{kerenidis2019q}  is a quantum alternative to the widely used $k$-means algorithm that provides a potential exponential speed-up in runtime while maintaining consistency with its classical counterpart.

Similarly, quantum algorithms for spectral clustering~\cite{kerenidis2021quantum}, a powerful technique for uncovering complex cluster structures, have been proposed, offering a polynomial speed-up compared to the classical runtime. Despite the progress and promises surrounding quantum unsupervised learning, several challenges remain. One of the primary limitations is the difficulty of loading and processing high-dimensional classical data into a quantum state, particularly in the absence of ideal amplitude encoding that could be enabled someday by quantum memory models like \emph{quantum random access memory} (qRAM)~\cite{PhysRevLett.100.160501}. Additionally, while PQCs show promise, they face training challenges such as barren plateaus, where the optimization landscape becomes flat and hinders training. Currently, they remain impractical for addressing significant classical tasks.
 
\medskip
\noindent\emph{Efficient quantum training of classical models}
\medskip

Quantum unsupervised learning~\cite{kerenidis2019q} utilizes quantum algorithms to identify patterns, clusters, or structures in unlabeled data. Clustering is a key methodology in unsupervised learning that groups data points into clusters based on their similarity. The goal is to partition data into groups where points within the same cluster are more similar to each other than to those in other clusters. Clustering typically works by identifying representative points, called centroids, for each cluster during training. Data points are iteratively assigned to the nearest centroid based on a predefined distance metric (e.g., Hamming or Euclidean).
Centroids are then recalculated to minimize the overall distance between points and their assigned centroids, repeating this process until convergence. Convergence occurs when cluster assignments or centroids stabilize, or when an objective function, such as the sum of squared distances, stops improving. This iterative process ensures that the algorithm identifies a stable grouping of data points that reflects the training data's structure. However, calculating these distances is computationally expensive, particularly in large data sets with a high number of features, as it requires evaluating all pairwise distances between data points.

Quantum clustering can address these challenges by leveraging the power of quantum algorithms, particularly in the computation of distances, which is at the core of clustering. One notable technique is the quantum swap test~\cite{buhrman2001quantum}, which enables efficient calculation of the similarity (or distance) between quantum states. By employing the swap test, a quantum computer can determine the inner product between two data points encoded as quantum states, allowing potentially for significantly faster similarity comparisons compared to classical methods. 

A prominent example of this is the acceleration of the $k$-means algorithm, where quantum methods can exponentially reduce the time complexity associated with calculating distances between data points and centroids by means of FTQC. Classical $k$-means involves iterative assignments of data points to clusters and recomputation of centroids, with each iteration requiring computationally expensive distance evaluations. A quantum variant performs these operations more efficiently. For instance, distance computations that scale linearly with the number of points and dimensions in classical approaches can scale logarithmically in quantum versions. Furthermore, the algorithm can encode cluster assignments in quantum states, enabling sampling and statistical analysis without directly revealing the entire data set. This quantum advantage would make unsupervised learning feasible for high-dimensional, large-scale data sets, offering significant speed-ups and enhanced scalability in real-world clustering applications under the assumption that the cost of encoding classical data into quantum states is negligible.

Importantly, while the calculation of the centroids can leverage quantum algorithms, one can also consider extracting this information from the quantum system and using a fully classical procedure to assign new points that were not originally part of the data set. This method is effective only if the training set is sufficiently representative and the number of new points is limited, ensuring that the computational cost of distance calculations remains feasible for classical resources. By integrating quantum techniques into the training process, clustering models can benefit from significant speed-ups in distance computation, making quantum clustering a powerful approach for tackling large-scale and high-dimensional data.

\medskip
\noindent\emph{Learning with quantum models}
\medskip

When adopting a quantum algorithm for both the training and testing phases of clustering, quantum methods can be leveraged to evaluate similarities, compute distances, and refine decision boundaries in high-dimensional spaces~\cite{lloyd2013quantum}. This approach is particularly beneficial when the training set is limited, and the amount of unseen data is significantly larger. However, its feasibility depends on the scalability of quantum resources and the cost of quantum-classical transitions.

A more structured form of unsupervised learning, image segmentation, also encounters major computational challenges when relying on classical methods. Supervised segmentation relies on large amounts of labeled data, but obtaining high-quality annotations is costly, time-consuming, and often inconsistent due to human subjectivity. In many domains, such as medical imaging and remote sensing, ground-truth labels are sparse, noisy, or unreliable, limiting the effectiveness of deep learning models trained on them. Unsupervised methods, including thresholding, clustering, region-growing, and edge detection, attempt to bypass the need for labels but come with their own limitations. These approaches often struggle with scalability, robustness, and adaptability to complex or high-dimensional data. More advanced graph-based methods, which represent images as graphs and segment them through optimization techniques like minimum cut or spectral clustering, offer a more structured approach but suffer from high computational complexity, particularly when solving large-scale combinatorial partitioning tasks.

Quantum computing presents a compelling alternative by leveraging quantum annealing to efficiently solve the combinatorial optimization problem underlying graph partitioning. This approach reformulates segmentation as a \emph{quadratic unconstrained binary optimization} (QUBO) problem~\cite{OptimizationReview}, allowing it to be executed on quantum annealers.

A key advantage is that in many cases, the grid structure of the graph representation of an image aligns well with the topology of quantum annealers, such as D-Wave’s Pegasus architecture. This enables efficient embedding, allowing the solution of QUBO problems with thousands of variables while maintaining a highly favorable ratio between logical binary variables and physical qubits.

By encoding pixels as nodes and similarities or dissimilarities as edge weights, quantum annealing can rapidly find optimal partitions, outperforming, in some circumstances, classical solvers in runtime, particularly for large and complex data sets. This capability is especially useful in scenarios with limited or noisy labeled data, where traditional supervised learning struggles. Furthermore, while current implementations focus on quantum annealing, the same segmentation framework can be extended to gate-based quantum computing, leveraging qubit-efficient variational quantum algorithms for further advancements in large-scale segmentation tasks. However, while quantum methods show empirical advantages on specific scenarios~\cite{PRXQuantum.3.030101}, their extensive practical adoption depends on continued improvements in quantum hardware scalability, noise reduction, and efficient quantum-classical hybrid workflows.

\subsection{Research directions}

\subsubsection{Learning models}

{\bf Focus on quantum-process-based learning models}: Most recent research in QML, particularly for near-term quantum systems, has centered on designing learning models around quantum processes. These include models such as quantum Boltzmann machines~\cite{Amin2018} and parametrized quantum circuits used as distribution generators (e.g., quantum circuit Born machines~\cite{Liu2018, Benedetti2019}) or as function approximators in supervised and reinforcement learning tasks. Considerable progress has been made in understanding the mathematical properties of these models, including their expressivity, complexity, and trainability bottlenecks.  Recently, efforts have also focused on identifying the boundaries between trainable and non-dequantizable models~\cite{gilfuster2025relation}, which contributes to a deeper understanding of when these quantum models can provide practical value. However, a critical question remains: why use these quantum models? The answer likely lies in future research, particularly with larger quantum systems where these models could demonstrate clear advantages.

{\bf Proving learning separations}: In parallel, a smaller but impactful line of research has concentrated on proving separations between classical and quantum learning, focusing on specific tasks where QML provides provable advantages. Initially developed for cryptographic tasks~\cite{PACLearning,QuantumFeatureEmbedding2}, these separations have been generalized to more complex problems in quantum many-body systems, establishing strong connections between QML and quantum complexity theory. The concept of ``data-hardness'' emerges here, referring to tasks where the complexity of the underlying correlations makes the problem accessible to quantum learning but not classical learning, even with data.  In other words, it can now be proven that there are problem domains where QML has advantages, even though these cases remain limited and abstract, in that they usually apply to highly structured data in order to make cryptographic proof tools available. The broader range of ML tasks whose efficiency cannot be mathematically characterized---which is the case for most of ML in practice---remains an area awaiting larger quantum system, and more NISQ-friendly methods. Also in such QML settings, questions of explainability occur naturally \cite{ExplainableQAI}.

{\bf Research motivation and objectives}: By combining these three key developments in QML, this research line aims to identify new domains of application where quantum learning can achieve significant breakthroughs, ideally in the near term. Specifically, the research line will focus on quantum methods for extracting useful features from hard-to-learn data sets, with applications in many-body physics and related computational tasks, 
in the cases they are likely ``data-hard.''
The ultimate goal is to develop quantum-assisted ML techniques that can outperform classical methods in certain well-defined domains of physics and beyond.\\

\noindent{\bf Short-term goals}
\begin{itemize}
    \item Identify classes of features extracted in the quantum phase. Develop an understanding of what makes certain features easier to extract using quantum machines.
    \item Target domains in many-body physics. Explore areas where quantum feature extraction could provide breakthroughs, such as:
\begin{itemize}
    \item Condensed-matter systems with highly correlated interactions and critical systems where classical ML methods fail.
    \item Exotic phases of matter, such as systems with topological order or symmetry-protected phases.
    \item \emph{High-energy physics} (HEP) applications, particularly in tasks related to quantum chromodynamics, where classical simulations are inadequate.
    \item Quantum control problems, where quantum feedback mechanisms could provide superior optimization capabilities.
\end{itemize}
\end{itemize}
\vspace{1em}
\noindent{\bf Mid-term goals}
\begin{itemize}
    \item Transition from toy models to real-world applications. Apply these methods to real-world problems in material sciences, solid-state physics, quantum chemistry, and HEP. The aim is to move beyond theoretical models and demonstrate the utility of quantum-assisted learning in practical domains.
\end{itemize}
\vspace{1em}
\noindent{\bf Long-term goals}
\begin{itemize}
    \item Optimizing material properties in quantum chemistry or solid-state physics.
    \item Anomaly detection: Analyzing HEP experiments data with hybrid quantum-classical models to extract novel insights or improve predictive accuracy.
\end{itemize}

\subsubsection{Quantum artificial intelligence for algorithmic discovery}

Quantum information processing bears the promise of solving some hard computational problems or of improving their ability to perform distributed tasks. While concrete examples of such algorithms and protocols have fostered the development of the field during the past 30 years, designing new ones is both a necessity and a challenge. Because there is no a priori criterion that will ensure the existence of an efficient quantum solution for a given task, quantum algorithmic development is akin to searching for a needle in a haystack. 

The goal here is to leverage quantum and classical AI to specifically address the issue of discovering and designing novel quantum algorithms and protocols. It involves utilizing quantum intelligent agents, specifically quantum neural networks, within a supervised and reinforcement learning framework. These agents will be trained to optimize quantum circuits that perform specific operations. Initial results have demonstrated that this approach produces optimal quantum circuits for tasks where quantum computers outperform classical ones, such as the \emph{quantum Fourier transform} (QFT)~\cite{AshleyOverview}, or nonlocal games (e.g., the CHSH game). In addition, they took into account hardware constraints, which suggest that these agents have not only the potential to uncover new, efficient quantum algorithms, a task that has proved to be a formidable challenge, but also to truly codesign.\\

\noindent{\bf Short-term goals}
\begin{itemize}
    \item Use the above quantum intelligence framework to re-discover known quantum algorithms and protocols (QFT, Grover’s search algorithm, CHSH game, quantum key distribution, etc).
\end{itemize}
\vspace{1em}
\noindent{\bf Mid-term goals}
\begin{itemize}
    \item Discover novel quantum algorithms and protocols, e.g., new algebraic transforms, cryptographic protocols, error correction codes, etc. 
    \item Discover circuits that inherently provide noise-robustness via suitable constraints during training, in an application-agnostic fashion, in order to use such circuits in the NISQ era.
\end{itemize}
\vspace{1em}
\noindent{\bf Long-term goals}
\begin{itemize}
    \item Use AI to co-design quantum algorithms and quantum hardware, optimizing the necessary resources and bringing forward the quantum applications era.
\end{itemize}

\subsubsection{Quantum data pre-processing}

Classical data-processing methodologies refer to the traditional approaches used in computing for the preparation and analysis of data. These methodologies encompass a variety of tasks, including, but not limited to, data cleaning, which involves the correction of errors or inaccuracies in the data set. Another key task is feature selection, which aims to determine the most significant variables for the analysis at hand.

Additionally, normalization processes are applied to adjust the data’s scale, enhancing the performance of models used for analysis. Techniques such as dimensionality reduction are also significant; they involve simplifying intricate data sets by reducing the total number of features while maintaining the data set's critical information. These methods are instrumental in ensuring that data is both accurate and pertinent, rendering it suitable for further analytical processes. This makes them indispensable tools in the realm of ML and data science.

Quantum data preprocessing builds upon classical techniques by using quantum computing capabilities. This process entails the encoding of classical data into quantum states, with the potential goal of enabling more efficient storage and manipulation of information. Furthermore, it involves the development of methodologies capable of processing quantum states to extract meaningful information. Quantum algorithms could achieve computational advantages in handling large-scale and high-dimensional data sets, where classical methods become computationally expensive or infeasible. One fundamental aspect of quantum data preprocessing is quantum state preparation, where classical data is mapped onto a quantum system using techniques such as amplitude encoding, basis encoding, angle encoding, and so on. These representations could enable parallel processing and efficient transformations, reducing the computational cost of subsequent analyses.

Quantum algorithms, such as quantum \emph{principal component analysis} (PCA)~\cite{QPA}, could allow for faster dimensionality reduction by using quantum linear algebra techniques. In particular, this approach offers an exponential speed-up over classical PCA by leveraging quantum density-matrix exponentiation. 
There are sparse sampling methods known with polynomial classical effort, even with a very high order of the polynomial. An exponential speed-up is possible only under stringent state preparation assumptions~\cite{QPA2}.
While classical PCA scales polynomially with system dimension, making it costly for large data sets, quantum PCA extracts principal components efficiently using multiple copies of a quantum system’s density matrix. This approach is especially beneficial for low-rank matrices, as it identifies dominant eigenvectors without requiring full matrix construction—provided a fault-tolerant quantum computer is available.

Similarly, quantum feature selection methods, utilizing variational quantum circuits and optimization techniques, may allow us to identify relevant variables more efficiently than their classical counterparts. Indeed, by encoding data into quantum states and optimizing parameterized quantum gates, these methods strive to explore complex feature interactions more efficiently than classical algorithms. Quantum parallelism enhances correlation detection and the efficient evaluation of multiple feature subsets, potentially leading to faster and more accurate feature selection.
Quantum distance metrics and kernel methods also play a crucial role in the preprocessing phase of QML by transforming data into a quantum-friendly format for further analysis. Quantum distance metrics, such as the quantum Hamming distance or the quantum Euclidean distance, can compute the similarity between data points by exploiting quantum parallelism. This enables the identification of subtle correlations in high-dimensional data that might be difficult to detect classically. Similarly, quantum kernel methods map data into a higher-dimensional quantum Hilbert space, where complex relationships are more easily separated, enhancing the ability to perform clustering and classification tasks.

Finally, quantum generative models, including quantum Boltzmann machines and quantum generative adversarial networks, present novel frameworks for data augmentation, consequently positioning them as valuable instruments within data preprocessing workflows. The integration of quantum computing into data preprocessing pipelines has the potential to unlock new possibilities for the efficient handling of large, complex data sets of various application domains.

\subsubsection{Quantum optimization}

Optimization is a fundamental problem in AI and beyond that involves identifying a point in a search space that minimizes or maximizes a given cost function. Many AI problems encounter extremely large search spaces, often due to combinatorial explosion in discrete domains, where the number of possible solutions grows exponentially or even faster. Additionally, in continuous search spaces, the lack of convexity in the cost function further complicates the process, leading to a potentially infinite number of solutions. This class of problems is often NP-hard in worst-case complexity.

Quantum computing offers three promising avenues to address these challenges~\cite{OptimizationReview}. First, quantum algorithms can encode and explore the search space of classical problems, potentially accelerating the discovery of optimal solutions, as envisioned with Grover's unstructured search algorithm in FTQC~\cite{grover1996fast}. For example, one can accelerate dynamic programming algorithms for the famous travelling-salesman problem in this way, although actual speed-ups have to be assessed for practical use cases compared to best-in-class classical solutions. This usually leads to polynomial quantum speed-ups. Second, efficient optimization techniques are essential for training quantum circuits in variational algorithms, with the promise of enabling more effective solutions to optimization problems. Here, no proven separations are known yet. 

Third, there are known problems where quantum algorithms can approximate instances of practically relevant problems such as integer linear programming well, while classical computers cannot approximate those instances well~\cite{OptimizationAdvantages}. In this sense, one can prove exponential separations of quantum over classical algorithms for sub-problems of hard optimization problems. Complementing this approach, decoded quantum interferometry~\cite{DecodedQuantumInterferometry} connects combinatorial optimization problems, like sparse max-XORSAT, to decoding local density parity check codes, a task that can be efficiently solved using classical algorithms such as belief propagation. This approach leads to a quantum algorithm that surpasses classical optimization techniques, including simulated annealing. This section explores the intersection of quantum computing and AI in both directions, highlighting their synergies in tackling optimization challenges.

{\bf Natively quantum gradient descent algorithms}: It is interesting to explore optimization algorithms finding approximate solutions efficiently that are natively suited to analog quantum simulators, as they can leverage the unique properties of quantum systems to address complex optimization challenges. For instance, \emph{quantum Hamiltonian descent} (QHD)~\cite{QuantumHamiltonianDescent} exemplifies this approach by mimicking gradient-based optimization through quantum dynamics. By utilizing quantum tunneling and other quantum-mechanical effects, QHD enhances the ability to escape local minima and navigate challenging optimization landscapes. Such methods highlight the potential of quantum-native algorithms to complement or outperform classical techniques in optimization, particularly in scenarios involving highly non-convex or high-dimensional problems.

{\bf Evolutionary optimization}: Evolutionary algorithms are search and optimization procedures inspired by the principles of natural selection and biological evolution. These algorithms emulate biological processes such as reproduction, mutation, crossover, and survival of the fittest, enabling the evolution of populations of candidate solutions over successive generations. In contrast to conventional optimization techniques, evolutionary algorithms operate on a population of solutions rather than a single point, thereby inherently facilitating parallel processing. This parallelism enables them to explore diverse regions of the search space concurrently, thus markedly enhancing their capacity to evade local optima and identify global solutions. Therefore, quantum computing is an appropriate means to improve the efficacy of evolutionary algorithms by capitalizing on its inherent parallelism and computational capacity. Quantum computers can speed up pivotal operations within evolutionary algorithms, including fitness evaluations, mutations, and crossover procedures, by processing superpositions of potential solutions in a concurrent manner~\cite{acampora2}. 


{\bf Quantum automated planning and scheduling} (QPS)~\cite{rieffel2015case} explores the integration of quantum computing into AI planning and scheduling tasks, focusing on \emph{quantum-supported planning} (QP) and \emph{scheduling} (QS). QP addresses tasks such as online planning, which involves real-time feedback, and offline planning, which is detached from execution. Under uncertainty, challenges such as \emph{partially observable Markov decision processes} (POMDPs)~\cite{barry2014QOMDP} arise, requiring complex strategies like conditional plans or utility-maximizing policies. These problems are computationally expensive, with tasks often being PSPACE-complete or worse. Initial quantum methods, such as quantum POMDP models and \emph{quantum Markov decision processes} (QMDPs), propose theoretical frameworks but lack concrete experimental validation.

Similarly, QS tackles problems like job-shop scheduling and its flexible variants (flexible job-shop scheduling), which aim to optimize job assignments on multi-purpose machines to minimize objectives like production makespan. Quantum approaches, including quantum genetic algorithms, quantum particle swarm optimization, and hybrid MILP-QUBO (mixed-integer linear programming with quadratic unconstrained binary optimization problem formulation) methods, demonstrate potential speed-ups in specific instances, such as solving scheduling tasks for up to a few hundred machines and jobs. Additionally, problems like bin packing have been addressed using quantum annealing, though current hardware limitations constrain their effectiveness. While early results highlight quantum advantages under specific conditions, further research is needed to identify scenarios where quantum methods outperform classical approaches.\\

\noindent{\bf Short-term goals}
\begin{itemize}
    \item Design and implement quantum-native optimization techniques and benchmark their performance against classical methods in AI applications such as combinatorial optimization and ML.
    \item Conduct medium-scale experiments to validate quantum-assisted scheduling and planning methods, identifying practical cases where quantum approaches offer computational advantages over classical techniques.
\end{itemize}
\vspace{1em} 
\noindent{\bf Mid-term goals}
\begin{itemize}
    \item Develop hybrid quantum-classical optimization frameworks and integrate them into industrial AI applications, such as logistics, supply-chain management, and financial modeling, leveraging near-term quantum devices.
    \item Design tailored quantum evolutionary algorithms optimized for specific quantum hardware and benchmark their performance on real-world optimization tasks, such as scheduling, bin packing, or combinatorial design, demonstrating practical advantages over classical approaches.
\end{itemize}
\vspace{1em} 
\noindent{\bf Long-term goals}
\begin{itemize}
\begin{samepage}
    \item Demonstrate a provable quantum advantage in practically relevant optimization tasks related to AI, such as large-scale automated planning, complex scheduling utilizing fault-tolerant quantum computers.
    \item Implement and validate quantum optimization techniques on large-scale industrial problems, using fault-tolerant quantum computers or specialized quantum hardware.
\end{samepage}
\end{itemize}

\subsubsection{Quantum reasoning}

Reasoning is the process by which intelligent agents draw conclusions, make decisions, and solve problems based on available knowledge, rules, and observations. It allows agents to infer new information from existing facts, resolve uncertainties, and adapt to changing conditions. Reasoning can be classified into several types, including deductive reasoning (deriving logically certain conclusions from general rules), inductive reasoning (inferring general principles from specific instances), and abductive reasoning (finding the most plausible explanation for given observations). Intelligent agents implement reasoning through formal logic systems, such as propositional logic, first-order logic, and probabilistic logic, to construct structured representations of their surroundings. They encode entities, relationships, and evolving conditions using knowledge representation techniques like ontologies, semantic networks, rule-based systems, and Bayesian networks.

These structured models enable agents to process complex information, infer new knowledge, and update their beliefs dynamically. Intelligent agents can analyze representations using inference mechanisms, which include logical inference engines, probabilistic reasoning models, and ML-based predictors. These mechanisms allow agents to generate new insights, optimal decisions, and adaptive action plans. As their environments evolve, intelligent agents continuously refine their models to maintain accuracy, coherence, and responsiveness in real time.

Quantum computing could enhance these capabilities by taking advantage of quantum parallelism to encode and process knowledge structures more efficiently. It could enable agents to evaluate multiple reasoning paths simultaneously, thereby significantly accelerating the processes of inference and decision-making.  Entanglement could enhance the representation of complex dependencies, improving context-aware reasoning and multi-agent coordination. Quantum interference has been shown to refine solutions by amplifying correct inferences and reducing computational errors.

The aforementioned quantum properties provide a fundamental advantage in fuzzy logic systems, probabilistic knowledge bases, and combinatorial optimization, thereby rendering AI agents more efficient in handling uncertainty, large-scale reasoning tasks, and highly dynamic environments. For example, in rule-based systems, quantum computing could provide the ability to evaluate multiple rules and potential conclusions simultaneously using quantum superposition~\cite{acampora1}. Unlike classical systems, which process rules sequentially or in parallel with significant computational overhead, a quantum circuit can encode an entire rule set and explore all possible inferences at once. This could drastically improve efficiency in expert systems, automated reasoning, and legal decision-making, where complex rule dependencies must be evaluated rapidly.

Additionally, quantum entanglement can enable richer representations of logical relationships, allowing for more nuanced and context-aware reasoning. Another goal of quantum computing in reasoning could be related to probabilistic inference and Bayesian networks~\cite{tucci}. Indeed, using quantum phase estimation and amplitude amplification, quantum algorithms can sample probability distributions potentially in a more efficient way than classical Monte Carlo methods. Moreover, other potential applications of quantum computing in reasoning are related to quantum-native reasoning architectures, where inference mechanisms exploit quantum entanglement and superposition rather than just mimicking classical logic. This could enable more advanced AI models capable of reasoning with fewer data and better than classical systems.

Another aspect of reasoning that quantum computing could significantly enhance pertains to the processing of natural languages. One of the approaches to \emph{quantum natural language processing} (QNLP) relies on category-theoretical representation of quantum mechanics and provides a framework for representing language meanings as quantum states and language operations as quantum processes \cite{1904.03478}.
 The high-dimensional Hilbert space can offer the modeling of complex semantic relationships, such as context-dependent meanings and interdependencies between words, in some cases achieving better results than conventional methods including large language models. An appealing feature of the approach is compositional generalization where the training involving smaller texts can be done even classically, while the application of the models on large texts requires the use of a quantum computer 
 \cite{2409.08777,EPTCS406.8}.

\noindent{\bf Short-term goals}
\begin{itemize}
    \item Develop quantum-assisted reasoning models to accelerate specific inference tasks (e.g., rule evaluation, probabilistic inference).
    \item Explore quantum-enhanced probabilistic reasoning techniques, such as quantum-assisted Bayesian networks and fuzzy logic systems.
    \item Implement proof-of-concept quantum circuits for evaluating multiple logical rules in parallel.
    \item Investigate hybrid quantum-classical approaches for automated reasoning and decision-making.
Benchmark quantum algorithms against classical methods in expert systems and legal decision-making.
\item Upscale experiments for QNLP for comparison against classical benchmarks.

\end{itemize}
\vspace{1em} 
\noindent{\bf Mid-term goals}
\begin{itemize}
    \item Optimize quantum reasoning architectures for scalability and efficiency in large-scale reasoning tasks.
    \item Integrate quantum-enhanced inference mechanisms into AI-driven decision-support systems.
    \item Develop quantum-inspired techniques for multi-agent reasoning and context-aware decision-making.
    \item Improve quantum algorithms for probabilistic inference, using phase estimation and amplitude amplification.
    \item Establish practical applications of quantum computing in real-world reasoning tasks, such as legal analysis, financial modeling, and medical diagnosis.
    \item Generalize QNLP to generative modeling.
\end{itemize}
\vspace{1em} 
\noindent{\bf Long-term goals}
\begin{itemize}
    \item Design fully quantum-native reasoning architectures that exploit entanglement and superposition for advanced AI cognition.
    \item Achieve quantum advantage in complex reasoning tasks, surpassing classical AI in efficiency and accuracy.
    \item Develop general-purpose quantum reasoning frameworks applicable across diverse AI domains.
    \item Explore quantum-enhanced learning models capable of reasoning with minimal data and handling extreme uncertainty.
    \item Integrate quantum reasoning with broader AI ecosystems, enabling next-generation autonomous systems with superior adaptability and intelligence.
\end{itemize}

\subsubsection{Quantum algorithms for multi-agent systems}

\emph{Quantum multi-agent systems} (QMAS) 
research explores the integration of quantum computing into the framework of autonomous agents and \emph{multi-agent systems} (MAS). This integration involves primarily designing quantum-enhanced methods for coordination and cooperation among agents in centralized and distributed environments. These systems consist of multiple agents, either homogeneous or heterogeneous, that collaborate or compete in complex environments to achieve both individual and joint goals simultaneously. Despite the maturity of classical multi-agent frameworks~\cite{weiss1999multiagent}, adapting these systems to leverage quantum computing is a relatively nascent field~\cite{klusch2004,klusch2007,hogg2007}. 

Quantum-supported coordination methods have been proposed to enhance agent collaboration. Notable advances include quantum coalition protocols and contract net systems, which are methods used to negotiate and form agent coalitions in competitive or cooperative settings~\cite{coal1,coal2}. Quantum coalition protocols have demonstrated reduced communication overhead and computational benefits over their classical counterparts. For example, quantum versions of coalition negotiation and resource allocation methods, such as quantum contract net protocols, offer enhanced privacy for agents while maintaining comparable computational efficiency~\cite{nesbigall}. 
However, these advantages are solely theoretical and require further exploration.

Beyond theoretical constructs, with the advent of near-term quantum technology, QMAS have begun to address real-world problems on existing quantum hardware. However, many proposed methods have been tested only on simplified models, and their scalability and utility for complex, real-world problems require further investigation. Nevertheless, the ongoing research highlights the potential of quantum computing to address computationally expensive challenges in multi-agent systems, offering a glimpse into the future of AI in quantum-enabled environments.

One key area where QMAS methods show promise is in enhancing cooperation and coordination among agents in a multi-agent system, particularly in complex optimization tasks such as resource allocation, network optimization, and social-network analysis. One fundamental challenge in coalition formation is to find a grouping of agents in coalitions such that their collective utility is maximized. Given the combinatorial nature of this coalition-structure-generation problem, that can be reformulated as a QUBO problem~\cite{Variational,OptimizationReview}, quantum annealing may be used to explore large solution spaces more efficiently. However, due to the current limitations of quantum hardware, scaling these approaches to larger problems remains an open challenge. 

Hybrid quantum-classical methods address these limitations by optimizing the interaction between quantum and classical resources. 
These methodologies delegate only the most quantum-suited tasks, such as solving graph-cutting problems, to quantum hardware, while classical components handle tasks that quantum systems cannot yet scale to. This strategic delegation results in a significant reduction in runtime and improved solution quality. Moreover, hybrid designs ensure that the advantages of quantum acceleration are leveraged without overburdening the limited qubit capacity of current hardware. 

Future research should focus on qubit-efficient variational quantum algorithms~\cite{tan2021qubit}, which further optimize quantum resource usage. These methods aim to minimize qubit requirements while maintaining quantum advantages in solving high-dimensional, combinatorial problems like coalition formation. By carefully integrating quantum and classical elements, hybrid quantum-classical approaches present a scalable path for addressing the computational challenges of coalition formation in MAS.\\

\noindent{\bf Short-term goals}
\begin{itemize}
    \item Develop qubit-efficient variational quantum algorithms to enhance coordination and decision-making in \emph{multi-agent systems} (MAS) while mitigating current hardware constraints.
    \item Experimentally validate quantum-enhanced multi-agent strategies, such as coordination and negotiation protocols, using near-term quantum devices and hybrid quantum-classical methods.
\end{itemize}
\vspace{1em}
\noindent{\bf Mid-term goals}
\begin{itemize}
\begin{samepage}
    \item Collaborate with hardware developers to improve quantum architectures tailored for MAS tasks, focusing on optimizing quantum circuits for agent interactions.
    \item Design more efficient integration strategies that balance quantum and classical computation to expand the scalability of QMAS solutions in 
    real-world AI-driven environments.
\end{samepage}
\end{itemize}
\vspace{1em}
\noindent{\bf Long-term goals}
\begin{itemize}
    \item Utilize FTQCs to enable entirely quantum-driven agent interactions, decision-making, and optimization, surpassing classical limitations in MAS applications.
    \item Deploy QMAS methodologies in high-impact domains such as autonomous systems, distributed AI, and dynamic resource management, leveraging advanced quantum hardware with improved stability and scalability.
\end{itemize}

\subsection{Use cases and applications}

\subsubsection{Healthcare and life sciences} 

As of 2025, the fields of medicine and life sciences face several significant roadblocks that hinder the full potential of medical advancements. One of the primary challenges is the high cost and complexity of developing new treatments, particularly in areas like gene therapy and personalized medicine. While gene therapy offers incredible promise for curing genetic disorders, issues such as targeting the right cells and controlling treatment dosage remain significant obstacles that need to be addressed before these therapies can become more widespread.

Another major roadblock is the integration of AI and ML into diagnostics and treatment planning. While AI has made significant strides in improving diagnostic accuracy, the technology still faces limitations in data quality and interpretability. Ensuring that AI systems can provide reliable, unbiased, and explainable results is crucial for their broader adoption in clinical settings.

By using ab initio simulations generated on quantum computers as training data sets for AI models, we could explore vast regions of chemical compound space that go beyond traditional bio-like molecules. This synergy between quantum computing and ML could significantly enhance drug discovery by providing a more comprehensive exploration of chemical spaces, identifying novel compounds, and accelerating the design process.

{\bf Quantum computing for accurate chemical simulations}: Quantum computers could accurately simulate molecular properties at the quantum level, which is critical in drug design, where small changes in molecular structure can drastically affect a drug’s efficacy. Classical computers struggle with simulating the behavior of large, complex molecules due to the exponential increase in computational demands, but quantum computers may handle these tasks efficiently, allowing for precise ab initio calculations.

{\bf Machine learning for efficient exploration}: While quantum computing may generate highly accurate data sets, it is currently limited in scalability. Machine learning models, on the other hand, excel in handling large data sets and exploring vast solution spaces efficiently. By training ML algorithms on quantum-generated data sets, we can develop models that generalize well to new, unexplored regions of the chemical compound space, predicting the properties of molecules that lie far from the bio-like compounds traditionally studied in drug design.

{\bf Expanding the search space}: Traditional drug discovery is often limited to molecules that are structurally similar to known bio-like compounds, which limits innovation. By leveraging ML models trained on quantum-derived data, we could explore chemical spaces that are far removed from this limited set, potentially uncovering entirely new classes of molecules with unique biological properties. This could lead to the discovery of novel therapeutics with mechanisms of action previously unknown to science.

{\bf Improving drug design}: Quantum simulations may provide insights into complex phenomena such as protein-ligand interactions, electronic structure, and reaction mechanisms with high precision. When these simulations are used as training data for ML algorithms, the resulting models can predict chemical properties like binding affinity, solubility, and reactivity more accurately. This reduces the time and resources needed for experimental testing and accelerates the identification of promising drug candidates.

\subsubsection{Industry} 

{\bf Image analysis}: Current state-of-the-art deep learning models for image analysis heavily rely on the availability of large and labeled data sets. However, with the rapid pace of data generation in fields such as medical imaging, autonomous driving, and satellite imagery, it is often unfeasible to label all available data due to the high costs, time, and expertise required for manual annotation. While unsupervised classical approaches, such as graph-based segmentation, provide alternatives for image processing without labeled data, they remain limited in practice because of the computational intensity required for handling high-resolution, real-world images. This challenge is especially evident in image-analysis tasks like segmentation and motion detection, where processing extensive data sets, complex patterns, and inherent noise or inconsistencies further complicates the computational workload. Quantum algorithms present a compelling solution, with the potential to reduce computational loads and increase efficiency in tasks like graph-cut optimization and classification. 

{\bf Safe navigation in autonomous driving}: The problem of \emph{collision-free navigation} 
(CFN) for self-driving cars is a complex optimization problem, often modeled as a POMDP. Current state-of-the-art solutions rely on deep reinforcement learning, which, while effective, requires substantial computing resources and training time. Quantum RL has emerged as a potential solution, showing faster convergence and improved stability in simplified environments. Quantum RL methods leverage quantum computation, specifically PQCs, which have demonstrated polynomial improvement in parameter space complexity compared to classical deep Q-networks. However, current QRL methods have not been tested in more complex real-world environments of CFN. However, initial research results show that leveraging quantum components, specifically for the critic in an actor-critic framework, offers potential advantages in training complex RL architectures with enhanced trainability and stability without requiring onboard quantum hardware during testing~\cite{sinha2025nav}.

{\bf Time-series analysis}: Time-series data, prevalent in domains like finance, healthcare, meteorology, and industrial monitoring, presents unique challenges for AI due to its temporal dependencies, high dimensionality, and potential for irregular sampling or noise. Classical methods, while powerful, often struggle to capture long-term dependencies or efficiently process massive data sets with intricate temporal patterns. Quantum methods, with their ability to process complex correlations and dynamics, offer promising advancements in time-series analysis. Quantum algorithms could enable more efficient modeling of temporal patterns, possibly using less memory than their classical counterparts, faster anomaly detection, and enhanced forecasting accuracy by leveraging the native ability of quantum systems to handle high-dimensional data and optimize over non-linear relationships. Additionally, quantum-enhanced versions of classical techniques, such as recurrent neural networks or transformers, might provide improvements in processing and predicting time-dependent data across a range of critical applications~\cite{Zhang2025}.

{\bf Bin packing}: Planning and scheduling problems are central to a wide range of industries, from logistics to manufacturing, and their complexity poses significant challenges as problem size grows. In logistics and supply-chain management, the bin packing problem is a prime example. This task involves efficiently packing items into the minimum number of containers without exceeding capacity limits, a critical issue in optimizing warehouse storage, transportation, and data center resource allocation. Quantum algorithms, particularly QUBO-based techniques, hold promise for solving these problems more effectively than classical methods, potentially improving space utilization and operational efficiency.

{\bf Job-shop scheduling}: In industrial production, the flexible job-shop scheduling problem is another key application. This task requires assigning jobs to machines to minimize production time, a complex challenge due to the vast number of potential job-machine combinations. Hybrid quantum-classical methods offer a practical approach by dividing the scheduling process into smaller, more manageable subtasks. These methods can enhance machine allocation, reduce production makespan, and adapt to dynamic industrial environments, making them valuable for advanced manufacturing systems in Industry 4.0. These applications demonstrate the potential of quantum technologies to revolutionize planning and scheduling in real-world settings.

{\bf Peer-to-peer energy trading}: Peer-to-peer energy trading is a complex problem typically framed using multi-agent systems, where the agents are individual energy producers (e.g., households with solar panels) and consumers negotiating energy exchanges. Each agent’s goal is to maximize utility, such as minimizing energy costs for consumers or maximizing revenue for producers, while ensuring the overall system remains balanced. The complexity arises from balancing supply and demand in real time despite the intermittent nature of renewable energy sources, designing fair and efficient pricing mechanisms, and optimizing infrastructure use, such as storage and grid connections. Additionally, regulatory and social considerations, such as compliance with policies and encouraging participation, add further challenges. Tailored quantum algorithms could offer advantages by optimizing dynamic multi-agent interactions, managing large-scale data for real-time decisions, and enhancing the efficiency of computationally intensive tasks like pricing and grid optimization.

{\bf Electric vehicle charging management}: The management of electric vehicle charging systems also relies on multi-agent systems, where the agents are vehicles, charging stations, and grid operators interacting to optimize energy distribution. Each agent has its own objective: the vehicles aim to minimize charging costs and waiting times, charging stations aim to maximize throughput, and grid operators aim to ensure grid stability and prevent overloading during peak demand. The primary challenges include scheduling charging sessions efficiently to prevent congestion, minimizing overall operational costs, and optimizing the geographical distribution of energy supply. The variability in electric vehicles arrival times, charging requirements, and station availability adds further complexity. Tailored quantum algorithms could enable efficient scheduling and routing optimization, improve real-time energy distribution, and balance the competing objectives of all agents while maintaining grid stability.

{\bf Dynamic environments in mobility and robotics}: a multi-agent system framework is crucial for coordinating interactions between diverse entities such as vehicles, drones, robots, and their environments. Each agent, whether an autonomous car or a robot, must optimize its decisions in real time to achieve goals like collision avoidance, efficient path planning, or task completion. Challenges arise from the dynamic and uncertain nature of environments, requiring agents to adapt to changes in traffic, obstacles, or tasks. Quantum algorithms for multi-agent systems could offer significant advantages by enabling faster optimization of complex interactions, improving decision-making under uncertainty, and scaling efficiently with the number of agents. Applications include optimizing coordinated fleet movements, enhancing task allocations in robotic swarms, and accelerating real-time computations for dynamic and unpredictable environments, paving the way for more robust and efficient autonomous systems.

\subsubsection{Quantum physics}

The rapidly evolving field of QML offers opportunities for tackling complex problems in physics, particularly in systems where classical ML techniques struggle, but also in other computationally hard tasks. Recent advances in QML highlight key research directions, which collectively point to new ways in which quantum computers could outperform classical systems for specific learning tasks, especially in complex quantum systems. In particular, the first proofs for quantum learning advantages from broad classes of genuinely quantum systems were provided in Refs.~\cite{molteni2024observables, gyurik2023separations}. Future research could build on these developments and explore the potential for quantum computers to extract valuable insights from challenging data sets in well-defined domains of physics, including many-body physics, condensed matter, and high-energy physics.

\section{AI for quantum}
\label{sec:ai4q}

\begin{figure*}
\centering
\includegraphics[width=0.8\textwidth]{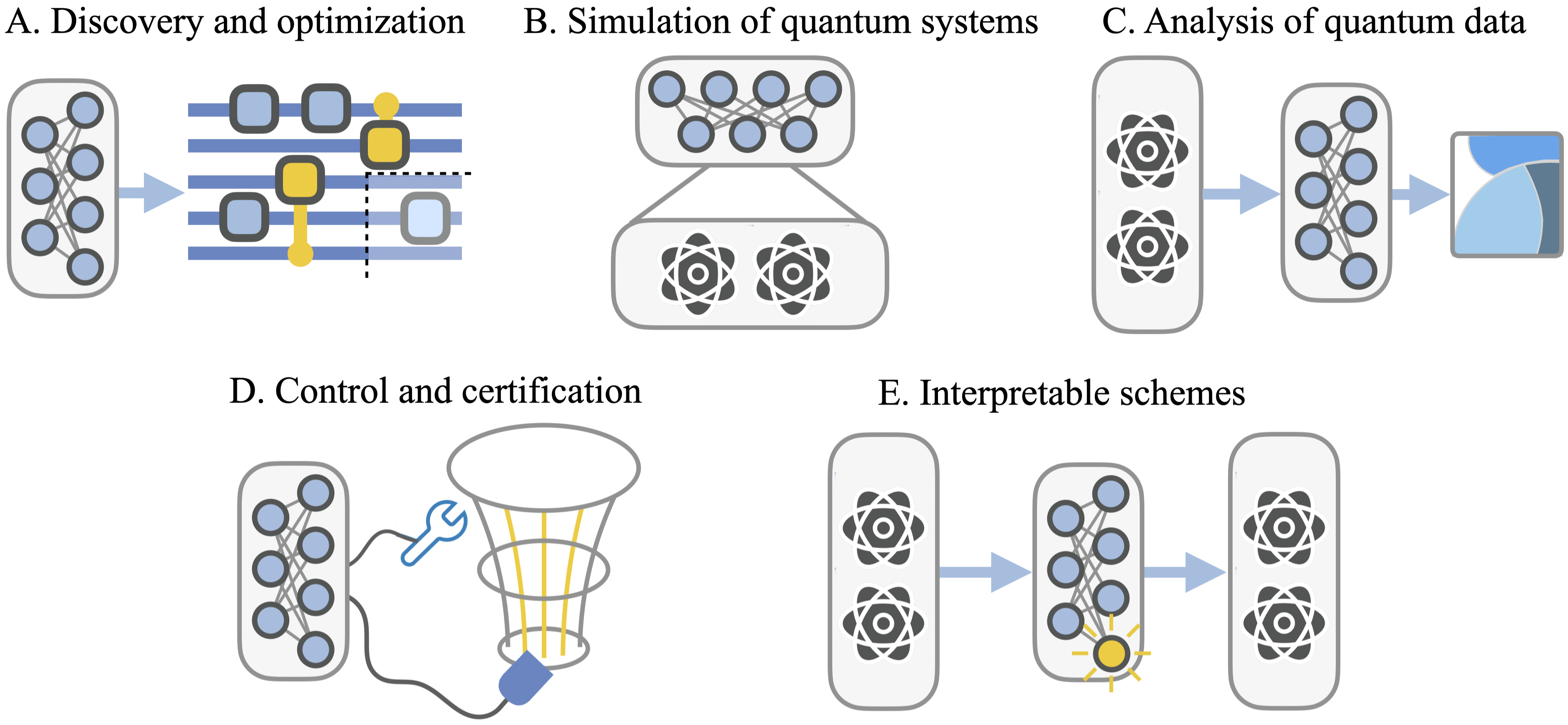}
\caption{Schematic of particular applications of AI for quantum computing, as presented in the different subsections of Section~\ref{sec:ai4q}. 
}
\label{fig:ai4quantum}
\end{figure*}

Artificial intelligence, ML, and computational methods inspired by them are transforming mathematical and computational modeling techniques across the board, offering in many cases a novel perspective on what it means to create a model and how to solve it. It accelerates work flows and moves the needle on the question of what can be simulated by classical means and even what intellectual task can be automated~\cite{RevModPhys.91.045002,DunjkoReview, krenn2023artificial,gebhart2023learning, alexeev2024}.

Quantum physics is a data- and computation-intensive science that presents several unique characteristics rooted in the basic principles of quantum mechanics. To begin with, our theoretical model fundamentally relies on a high-dimensional object --- the quantum state. Whenever handling it in classical terms, we have to resort to high-dimensional data such as samples of measurement outcomes of experiments or state-vector representations in numerical simulation. Moreover, quantum measurements are fundamentally probabilistic and inevitably entail back-action on the quantum state. Finally, entanglement constitutes a key distinctive feature of quantum systems. 

Quantum computers and quantum simulators allow us to explore these quantum characteristics and their consequences in unprecedented detail. But at the same time, exhausting these possibilities and the further advancement of experimental and theoretical techniques poses new challenges. For example, the highly resolved observations call for new approaches for data analysis and the possibilities for precise control demand finding optimal strategies in high-dimensional control spaces. Moreover, the strong precision requirement of quantum technology requires testing against complex, multi-parameter models taking the tiniest disturbances into account. 

With its known strengths in pattern recognition, dimensional reduction, and strategy discovery, AI provides valuable new tools to address central challenges in quantum physics related to high-dimensional data, precise manipulation, and modeling. Therefore, AI can become a key enabler in quantum science with impact ranging from fundamental insight to technological applications.



\subsection{Discovery of quantum hardware and algorithms}\label{sec:discover_hardware}

The routine operation of quantum computers will require new approaches for optimal control~\cite{niu2019universal, bukov2018reinforcement, Baum2021, vaidhyanathan2024, berritta2024PRApp}, readout~\cite{seif2018machine, liu2020repetitive, Lienhard2022, hickie2024}, efficient compilation~\cite{fosel2021quantum, Pozzi2022, furrutter2024quantum, moro2021quantum}, and error correction and decoding~\cite{fosel2018reinforcement,olle2024simultaneous, nautrup2019optimizing, sivak2023real,Reinforcement,andreasson2018quantum,PhysRevLett.131.050601}, or entire de-novo design strategies of experiments and devices~\cite{krenn2016automated,knott2016search,Melnikov2018,krenn2020computer,menke2021automated,ruiz2023digital,landgraf2025automated} -- some of which even benefit from quantum computers themselves~\cite{kyaw2021quantum}. One example is ML models for compilation that can autonomously learn generic strategies to compress quantum circuits~\cite{fosel2021quantum, ostaszewski2021reinforcement}. Another example involves the design of new quantum-enhanced measurement devices with improved sensitivity, that can act as diagnosis tools for quantum computers~\cite{o2019hybrid,huang2025quantum}.


This illustrates that the ML toolbox provides manifold ingredients that can boost the hybrid quantum-classical operation for these purposes. In addition, ML is one of the most prominent tools for an efficient interface between classical and quantum systems --- both for experiment discovery and control as well as for the translation of classical data into a quantum state and the transpilation of classical into quantum algorithms. Machine learning algorithms operating in hybrid classical-quantum hardware will also allow us to harness properties of quantum systems to devise energy-efficient control tasks.


\subsubsection{Quantum architecture search with machine learning for near-term algorithms}

\emph{Variational quantum algorithms} (VQAs)~\cite{Variational} are widely used in the NISQ era to solve ML problems such as classification, prediction, and generative tasks. The main component of VQAs is parameterized quantum circuits, but such circuits encounter trainability issues, such as barren plateaus~\cite{mcclean2018barren, Larocca2025}. 
Quantum hardware constraints and noise further worsen the performance of VQAs. To address these issues, there is a need to design efficient circuits with an optimal set of parameters tailored to underlying problems and quantum hardware; such a task of generating circuits for a specific problem is also referred to as quantum architecture search or automatic design of circuits.

In the early 2000s, researchers used evolutionary algorithms to find efficient quantum circuits. Since then, various ML and optimization techniques such as deep reinforcement learning, Bayesian optimization, adaptive methods such as quantum autoencoders for data compression, other attention-based models, or reinforcement learning are used to generate optimal circuits~\cite{krenn2023artificial}. Many techniques for quantum architecture search are inspired by neural architecture search and can be used to generate an efficient ansatz, which can help in tackling issues associated with VQAs, co-designing algorithms for specific quantum hardware, or finding new efficient and better algorithms.
The emerging capabilites of quantum processors to include feedforward components that allow conditioning subsequent operations on measurement outcomes, particularly call for leveraging the strengths of ML algorithms in strategy discovery~\cite{Iqbal23,porotti2023gradient,Alam2024,Puente2025}.

\subsubsection{Compilation of quantum circuits}

Many quantum algorithms are first designed on an abstract level, where it is assumed that one has access to many different quantum gates and that any qubit can be connected to any other. However, actual quantum hardware generally has a quite limited (but universal) gate set available, and usually only has a limited nearest-neighbor connectivity in some two-dimensional qubit layout. Therefore, most quantum circuits designed to implement a quantum algorithm need to be compiled (transpiled) to fit on specific quantum hardware. This compilation has several parts, e.g., efficient conversion of gates in the algorithm to the available native gate set, initial mapping (allocation) of logical qubits from the abstract algorithm to physical qubits on the hardware, ``routing'' of qubits that need to interact in the algorithm across the hardware by inserting gates for swapping to make up for the limited connectivity, and compression of the circuit through merging gates using circuit identities~\cite{Siraichi2018, Paler2021, Yan2024}. The aim of the compilation is to reduce the circuit depth and resources required to run a quantum circuit, to minimize the impact of errors and maximize the reliability of the circuit output. These compilation problems, which tend to NP-hard~\cite{Siraichi2018, Ito2023}, also extend to distributed and error-corrected quantum computing.

Due to the hardness of compilation, exact or close to exact methods can become too time-consuming already for tens or hundreds of qubits. Therefore, heuristic approaches tend to be used~\cite{Li2019, Sivarajah2021}, including methods involving a category-theoretic representation of quantum circuits such as the familiar \emph{ZX calculus} 
\cite{2012.13966,Coecke2007graphicalcalculus}.
Nonetheless, there is increasing interest in improving on heuristics using ML~\cite{Herbert2018, fosel2021quantum, ostaszewski2021reinforcement, Pozzi2022, Sinha2022, Zhou2023,2312.11597,QCTRL1}. These ML approaches tend to be based on reinforcement learning to find better strategies for playing the ``game'' of compilation, where a ``move'' consists of placing a certain gate or swapping the routing in the quantum circuit. Indeed, recent works build on the well known RL algorithm AlphaZero for qubit routing~\cite{Tang2024} and compilation for fault-tolerant quantum computers~\cite{Ruiz2024}. 
In this same line, it is also feasible to make use of AI to identify notions of error-robust quantum logic optimization \cite{QCTRL3}.
Recently, generative approaches have also been proposed for circuit generation~\cite{furrutter2024quantum}, circumventing the need of RL approaches to simulate at each step their ``move'' and the resulting circuit, a task that is computationally very expensive.

\subsubsection{AI for quantum sensing}

Similar uses of AI to enhance and optimize quantum protocols have been studied in the context of quantum sensing, a field of research dedicated to achieve extremely precise and sensitive measurements of physical quantities. It exploits quantum phenomena such as superposition and entanglement to surpass the limitations of classical sensors, presumably enabling advancements in various 
fields \cite{RevModPhys.89.035002}. In this context, finding appropriate probe states can be excessively difficult, in particular since
due to a subtle interplay between properties of entanglement and 
concomitant robustness to noise, one cannot simply optimize a simple figure of merit for perfect quantum circuits. For this challenging task, the use of AI has once again been suggested \cite{krenn2023artificial,QCTRL2,AISensing,AISensing2}.

\subsubsection{AI for quantum error correction}

Many technologically or scientifically relevant applications of quantum computers require the execution of deep quantum circuits. Although progress is made by developing more efficient algorithms and using error mitigation, to achieve the full potential of quantum computing, qubit error rates (measured as typical gate time over qubit coherence time) should be reduced from current best values of $p\sim
10^{-3}$ to $p\sim 10^{-10}$ or lower. The currently most promising framework to achieve this is to use topological stabilizer codes~\cite{RevModPhys.87.307}, such as the surface code~\cite{Fowler2012}, where logical qubits consist of entangled physical qubits. Such codes require a classical control algorithm, a ``decoder'', that implements error correction based on information which is fundamentally limited by its quantum nature. Traditional decoders are based on sophisticated algorithms (such as minimum-weight perfect matching) that require detailed knowledge of qubit decoherence during gate operations, as well as measure and reset errors. In contrast, ML-based decoders can be trained on simulated data, then fine-tuned on experimental data to provide device-optimized performance, while avoiding the intrinsic limitations of non-trainable algorithm-based decoders. Even though such decoders, using supervised learning, have recently been shown to outperform state-of-the-art decoders in accuracy for a surface code~\cite{Lange2023data,PhysRevResearch.7.013029,bausch2024learning}, challenges remain with scaling up to large code distances, to generalize to other stabilizer codes, and to be able use them for real-time error correction, which requires $\mu$s decoding times for superconducting quantum computers.

Another area where AI may have a significant impact on quantum error correction is in finding new error-correcting codes, where an important measure for future implementations in hardware is the code ratio, i.e., the fraction of logical qubits to physical qubits, as well as the feasibility of implementing stabilizer measurements and logical gates. For these types 
of problems, involving discovery of new algorithms, reinforcement learning shows great promise~\cite{fosel2018reinforcement,nautrup2019optimizing,olle2024simultaneous}.\\

\noindent{\bf Short-term goals}
\begin{itemize}
    \item AI designs of new experimental setups able to create target quantum states, design of experimental quantum devices with superior sensitivity, quantum circuits with optimized gates and hardware constraints.
    \item AI-based quantum error decoders extended to large code distances and beyond the surface code.    \item 
    \item Improved design of probe states in quantum sensing.
\end{itemize}
\vspace{1em}
\noindent{\bf Mid-term goals}
\begin{itemize}
    \item Discovery of more efficient implementations of known quantum algorithms as well as better ansatzes, e.g., for quantum chemistry and quantum simulation problems.
    \item Generate optimal circuits and new quantum algorithms for industrial use cases or applications.
    \item AI-based compiling of quantum circuits with hundreds of qubits outperforming non-AI approaches. 
    \item AI-based quantum error decoders for real-time decoding.
    \item AI-based discovery of novel high code-ratio codes with practical stabilizer circuits and efficient gate implementations.
\end{itemize}
\vspace{1em}
\noindent{\bf Long-term goals}
\begin{itemize}
    \item End-to-end generation of quantum routines for input tasks, from devising new quantum algorithms or variational approaches to their efficient implementation in the native gates of the given hardware.
    \item Use AI techniques to discover new strategies (for compilation of quantum circuits, quantum error correction, etc.) and better quantum algorithms.    
    \item Practical fault-tolerant quantum computing controlled by AI-based classical infrastructure.    
\end{itemize}

\subsection{Simulation of quantum systems}

The simulation of quantum many-particle systems on classical computers is one of the greatest challenges in physics~\cite{Georgescu2014}. The exponential growth of the underlying Hilbert space with the number of constituents limits full state-vector simulations to systems of no more than around fifty qubits, if no further structure can be exploited. The sophisticated toolbox of quantum many-body physics and quantum chemistry exploits specific properties of the quantum states of interest for efficiency, allowing to reach far beyond the capabilities of full state-vector simulations in various problem families. However, many relevant cases remain hard, especially in the presence of strong entanglement (or ``magic''). Precise simulations furthermore rely on accurate models of the system of interest. In quantum computing, the detailed understanding of the underlying physical platform is of utmost importance for the precise manipulation of the system required to accomplish computational tasks. Finding accurate descriptions is, however, a major challenge, as the relevant degrees of freedom may interact in manifold ways and many environmental effects may play a role.

In view of quantum computing, numerical simulation capabilities are relevant from two points of view. As already alluded to above, quantum many-body simulations are key for the design and certification of quantum devices. Besides that, the state of the art in numerical simulation defines a bar for quantum advantage. As quantum simulation for physics and chemistry is among the most promising use cases of quantum computing, pushing classical simulation methods targeted at quantum many-body systems in general is an important objective for the honest assessment of quantum advantage. 

AI methods will play a central role for advancing the numerical modeling of complex quantum systems, as we outline in the following.

\subsubsection{AI-boosted numerical simulation}
\label{sec:numerical_simulation}

Finding innovative ways of dealing with the quantum curse of dimensionality becomes inevitable when correlations come into play. AI building blocks --- particularly, ML --- are increasingly entering numerical simulation algorithms to solve central computational challenges. In \emph{density functional theory} (DFT) for ab initio quantum physics and chemistry, correlation effects are captured by a generally unknown exchange-correlation functional and finding suitable forms of this functional is key for accuracy. Using ML techniques to solve this problem at the heart of DFT has been identified as a promising avenue~\cite{Snyder2012,Dick2020,Fiedler2022}. Along a slightly different line, ML models can be trained to predict beyond-DFT corrections and accelerate DFT-based molecular dynamics simulations~\cite{Bogojeski2020}.

When targeting the properties of strongly correlated systems, Monte Carlo methods constitute a powerful and widely used approach. As they typically employ sampling by generating Markov chains, many Monte Carlo methods crucially rely on suitable update policies to efficiently explore the relevant configuration space. However, finding such policies can become very difficult in regions of particular interest, such as phase transitions, where critical slowing down prohibitively affects naive schemes. ML-based generative models have been proposed as a novel route for efficient proposal strategies, applicable across various flavors of the Monte Carlo method, including both low- and high-energy physics~\cite{Liu2017,Huang2017,Xu2017,Song2019,Albergo2019}. In a similar spirit, ML models can be trained to generate typical low-energy states in a one-shot fashion~\cite{Noe2019, Zhao2019}, or --- going a step further --- to learn the full distribution by variational minimization of free energy~\cite{Wu2019}.

Efficient solvers for the quantum many-body problem may be limited to yielding Green's functions in imaginary time, such that extracting physical information requires analytic continuation. Solving this notoriously ill-posed inverse problem is an outstanding challenge in the field, for which ML approaches have been suggested as a way forward~\cite{Yoon2018}.
In view of wave-function-based methods, neural-network representations of the many-body state have been introduced as a new paradigm~\cite{carleo2017solving}, which will be outlined in more detail below.

Overall, ML techniques are entering the entire spectrum of numerical simulation methods as enablers that mitigate or resolve central bottlenecks. 
Similar to the paradigmatic shift in the area of protein folding~\cite{Jumper2021}, it can be anticipated that ML methods will lift computational approaches for complex quantum systems to a new level.
While current work focuses on tackling specific components of existing methods, leveraging capabilities of AI may also be a way forward to address the outstanding challenge of simulation across the whole range of length or energy scales, that today cannot be covered in a unified manner.

\subsubsection{Neural quantum states}

\emph{Neural quantum states} (NQS) constitute a new versatile class of variational wave functions~\cite{carleo2017solving,Lange2024,Medvidovic2024}. The underlying idea is that the proven capabilities of neural networks in pattern recognition and generalization make them well-suited for the compressed classical representation of quantum states. Since neural networks are universal function approximators in the limit of large network sizes, arbitrary wave functions can be represented as NQS in principle. While the ultimate limitations of NQS are not yet known, it has been proven that NQS cover a strictly larger part of the Hilbert space than the tractable tensor network states --- the main alternative~\cite{Sharir2022}. Therefore, it is expected that NQS-based approaches can substantially push the limits of numerical simulation in particularly challenging cases, such as two- or three-dimensional lattice systems, ab initio simulations, or in the presence of strong entanglement. 

In pioneering works, the efficiency of simulating dynamical problems~\cite{Schmitt2020,Reh2021,Nys2024} --- including quantum circuits~\cite{Medvidovic2021} --- and of addressing low-energy properties of molecular systems~\cite{Hermann2023} as well as exotic states of matter~\cite{robledo_moreno_2022,Chen2024} have been demonstrated.
Recent research explores the possibilities of NQS foundation models to simultaneously solve whole families of physical systems~\cite{Rende2025} or hybrid simulation approaches incorporating data from a quantum simulator (or computer) to profit synergetically from the complementary strengths~\cite{Czischek2022}.

Neural quantum states can be expected to have a transformative impact over a range of fields, where the quantum physics of interacting particles is of interest. These fields include condensed matter, nuclear, and high-energy physics, quantum chemistry, and --- specifically in view of quantum technologies --- emulation of quantum devices. Being still at early stages of development right now, its versatility will allow the technique to become an established part of the computational toolbox. Important steps on the way will be the development of a solid theoretical underpinning to delineate the range of applicability, an encompassing toolbox for incorporating physical bias where needed, and unified optimization algorithms without the need for case-by-case adaptation. Efficient implementations of NQS algorithms provided in open-source libraries will foster wider use of the approach and, ultimately, synergetic operation is conceivable in conjunction with quantum information processors.

\subsubsection{Surrogate models and digital twins}

The simulation techniques discussed above generally assume the description of the system of interest in terms of a model rooted in the underlying physical theory. Using ML methods, one may instead learn new models describing the physics of interest, for example, in the form of artificial neural networks. Some of the ML methods for enhanced Monte Carlo sampling discussed above in Section~\ref{sec:numerical_simulation} are, in fact, an example of that.
More generally, AI-based surrogate models can be a way to bypass expensive simulations by directly predicting physical properties of a system for given input parameters. 
Surrogate modeling of the dynamics of quantum systems may for example enhance the efficiency of pulse engineering for quantum control~\cite{Mohseni2022}, and ML models of interatomic potentials or force fields can replace demanding ab initio computations in molecular dynamics simulations~\cite{Bartok2017,Fiedler2022}.
Beyond ``black-box'' models, different degrees of interpretability can be imposed to learn effective models of correlated quantum systems~\cite{Rigo2020} or for Hamiltonian learning, which is highly relevant for quantum device certification~\cite{gebhart2023learning, Craig2024}. 

By learning highly accurate models of quantum systems, these techniques lay the foundations for creating digital twins of quantum devices.
Widespread application will, however, rely on faithful uncertainty quantification and a high degree of interpretability while maintaining generality of the models.
\\

\noindent{\bf Short-term goals}
\begin{itemize}
    \item Identification of promising use cases, where AI ingredients fundamentally advance numerical simulation.
    \item Incorporation of physical bias (e.g., symmetries) into ML building blocks.
    \item High-efficiency emulation of quantum devices for quantum control.
    \item Demonstration of advantage of AI-enhanced simulation in new applications.
\end{itemize}
\vspace{1em} 
\noindent{\bf Mid-term goals}
\begin{itemize}
    \item Unified frameworks for AI-boosted simulation methods.
    \item AI-supported interoperability of numerical methods.
    \item Foundation models for simulation of quantum systems.
    \item Hybrid AI-assisted quantum-classical simulation approaches.
    \item Encompassing theoretical understanding of the scope and limitations of AI-enhanced simulation of quantum systems.
    \item Faithful uncertainty quantification and interpretability for simulation pipelines involving AI methods.
\end{itemize}
\vspace{1em} 
\noindent{\bf Long-term goals}
\begin{itemize}
    \item A bar for quantum advantage in simulating quantum systems at the limits of classical capacity in the age of AI.
    \item End-to-end AI-based models of quantum devices.
    \item Unified simulation of complex quantum systems across length and energy scales.
\end{itemize}

\subsection{Analysis of quantum data}

Data from quantum experiments is particularly complex and hard to analyze for various reasons. 
The high dimensionality of the state space of quantum many-body systems and the intrinsic property of projective measurements to reveal only partial information about the system state make the characterization of quantum states and processes a notoriously hard task. 
Furthermore, long experimental cycle times often severely limit the available measurement statistics, and readout may be restricted and suffer from technical noise.
Machine learning methods have been shown to be a powerful tool for ameliorating these challenges by means of, e.g., dimensional reduction, feature extraction, generative modeling of data distributions, or learning adaptive measurement strategies.


\subsubsection{Quantum state and process tomography}

For many applications in quantum computing, sensing, and communication, it is necessary to prepare quantum states with specific properties and to certify the successful preparation of these states by means of full state reconstruction.
This task becomes extremely challenging for multiqubit systems due to the high dimensionality of the reconstructed states.
Incorporating prior knowledge about the reconstructed state, such as knowledge about the purity or symmetries, helps to reduce the complexity and required experimental resources.
Here generative modeling provides a general ansatz for augmenting the statistical confidence by exploiting structure in the sampled data distributions~\cite{gebhart2023learning, Torlai2018}. Examples of ML methods used in this field include conditional generative adversarial networks~\cite{Ahmed2021}, convolutional neural networks~\cite{Schmale2022}, attention-based tomography~\cite{Cha2022}, and recurrent neural networks; ML-inspired gradient-descent methods have also been proposed~\cite{Gaikwad2025}.

Such methods can also be applied to quantum process tomography, the full characterization of quantum operations, which is crucial for characterizing and optimizing quantum gates, for example. Unsupervised learning~\cite{Torlai2023}, neural-network digital twins for error mitigation~\cite{Huang2025}, and ML-inspired stochastic gradient-descent methods~\cite{Ahmed2023} are some of the approaches that have been used here. 
In the context of quantum-simulation experiments, one is interested in certifying the correctness of the emulated Hamiltonian dynamics, which can be achieved by Hamiltonian learning techniques~\cite{hangleiter_robustly_2024}. 
In addition, active learning can be used to find optimal adaptive measurement strategies~\cite{Quek2021, Lange2023}. 
Here, given a set of previous measurements, one aims to determine the optimal next measurement setting that maximizes the expected information gain about the quantum state or process.



\subsubsection{Feature extraction and dimensional reduction}

The extraction of physical properties of interest from measurements in quantum many-body systems is often impeded by the high dimensionality of the data distributions or imperfect and noisy readout.
Here, ML methods can be exploited in various ways. 
First, denoising methods can be used for processing experimental images, for example, for the reliable determination of atom numbers in quantum gas microscopy experiments limited by photon shot noise and finite imaging resolution~\cite{Impertro2023}.
Second, the dimensionality of data can be reduced to extract relevant physical features by means of interpretable and explainable ML models, as we further review in Section~\ref{sec:ae}. 
This can result in the unbiased identification of the essential observables, which is particularly important in noisy settings or for exotic order, where traditional methods fail.
Third, supervised-learning methods can be used for classifying phases of matter, including cases where the relevant order parameter is not known a priori~\cite{carrasquilla2017machine}. 
As a final example, the quality of approximate descriptions of physical processes has been accessed by testing the ability of a neural network to distinguish model predictions from real experimental data~\cite{bohrdt2019classifying}.

\subsubsection{Quantum error mitigation via post-processing by AI}

Quantum algorithms typically generate probability distributions over possible measurement outcomes. These distributions encode the solutions to computational problems, such as optimization, simulation, or factorization. However, due to inherent hardware limitations, noise, and decoherence, the observed probability distributions often deviate significantly from their ideal counterparts. This discrepancy poses a major challenge for practical quantum computation, particularly in the current era of NISQ devices.

From a computational perspective, errors in quantum probability distributions emerge due to several factors. These include gate imperfections, state preparation and measurement errors, and decoherence caused by the environment. These errors distort the statistical properties of quantum algorithms, resulting in unreliable outputs. Unlike classical errors, which can often be corrected deterministically through redundancy, quantum errors require sophisticated mitigation techniques~\cite{Cai2023} due to the no-cloning theorem and the fragility of quantum states. There are also worst~case bounds known that limit the applicability of quantum error mitigation methods to basically log-log-depth quantum circuits in worst case complexity.

Recent advancements in AI techniques, particularly ML, reasoning, and optimization, have emerged as powerful tools to correct and refine quantum probability distributions post-processing. Machine learning models, such as neural networks and Gaussian processes, have the capacity to learn error patterns from experimental quantum data and apply corrections to infer the ideal distribution~\cite{kim2020quantum}. Furthermore, reinforcement learning and Bayesian reasoning can enhance error mitigation by dynamically adapting correction strategies based on observed system behavior. 
Approximate reasoning, as exemplified by fuzzy logic, plays a pivotal role in the mitigation of quantum errors by addressing the inherent uncertainty and imprecision in the identification of quantum error patterns~\cite{acampora3}. 
%
Furthermore, optimization methods such as genetic or evolutionary algorithms enable the precise adjustment of post-processing parameters to restore fidelity to the quantum outputs~\cite{acampora4}. 

The utilization of AI-driven techniques has emerged as a promising approach to enhance the reliability of near-term quantum computations. This enhancement is achieved without necessitating additional quantum resources, thus demonstrating the efficacy of AI-driven techniques in quantum error mitigation.
\\


\noindent{\bf Short-term goals}
\begin{itemize}
    \item Develop supervised and unsupervised learning methods that can characterize data arising from various types of quantum hardware.
    \item Introduce interpretable and explainable models and architectures able to rediscover known quantum theories and phenomena directly from data.
    \item Develop ML models to learn and correct quantum error patterns from experimental data.
    \item Implement fuzzy logic-based approaches for identifying and mitigating quantum errors.
    \item Apply evolutionary optimization techniques to refine error-mitigation strategies.
    \item Benchmark AI-driven post-processing methods against traditional quantum error mitigation techniques.
\end{itemize}
\vspace{1em}
\noindent{\bf Mid-term goals}
\begin{itemize}
    \item Build autonomous pipelines able to create, directly from data, new theories and efficient descriptions of the data arising from quantum processes.
    \item Develop AI-assisted dynamic parameter tuning for post-processing quantum outputs.
    \item Investigate scalable AI techniques to handle characterization of increasingly large quantum systems.
    \item Integrate AI-driven quantum error mitigation into practical NISQ applications, such as optimization and simulation.
\end{itemize}
\vspace{1em}
\noindent{\bf Long-term goals}
\begin{itemize}
    \item Combine automated methods for processing of quantum data with those proposing and controlling quantum systems to create better experiments from which to acquire new theoretical insights.
    \item Achieve AI-enhanced quantum error mitigation techniques that enable practical quantum advantage.
    \item Establish AI-quantum co-optimization strategies for next-generation quantum processors.
\end{itemize}

\subsection{Automated control and calibration of quantum technologies}
\label{sec:automatedcontrol}

Both quantum technologies and fundamental experiments rely on the precise device fabrication and control of a large number of quantum degrees of freedom. Common challenges are the efficient characterization~\cite{gebhart2023learning, hickie2024, straaten2022, severin2024, Schuff2023}, avoidance, and mitigation of inevitable hardware imperfections~\cite{Craig2024}, data availability~\cite{zwolak2024}, and optimization of control strategies~\cite{Rao2025, Baum2021, NamNguyen2024, huang2024, berritta2024realtime}. As the qubit numbers grow and the applications become manifold, automation becomes essential and ML tools have been shown to be powerful for this purpose~\cite{Ares2021}.

Examples include the fully automated tuning of quantum-dot devices~\cite{schuff2024fully} and the automated optimization of entangling operations on a superconducting quantum processor~\cite{cao2024}, which could substantially improve their quality, crucial for near-term applications. ML will accelerate the process of quantum state preparation, gate operations, and measurement, leading to faster and more efficient quantum computation or quantum communication protocols on given hardware~\cite{alexeev2024}. Thus, the ML toolbox can be employed to extend the applicability of quantum devices to problems with many noisy parameters such as imaging, radar, or gravitational wave detection. ML methods can readily be utilised to characterize noise sources~\cite{berritta2024PRApp, Craig2024}. 
\\


\noindent{\bf Short-term goals}
\begin{samepage}
\begin{itemize}
    \item Online control of quantum experiments via reinforcement learning and other methods, improving the stability and coherence of quantum systems.
    \item Development of better state preparations, operations and measurement patterns, enhancing the quality of current experiments.
\end{itemize}
\end{samepage}
\vspace{1em} 
\noindent{\bf Mid-term goals}
\begin{itemize}
    \item Development of fully-automated quantum laboratories, easing the preparation of multi-component experiments (e.g., calibration of optical tables and similar), for a wide variety of given tasks (i.e., multi-task ML agents). This would be combined with \emph{large language models} (LLM) for easy user interaction.
\end{itemize}
\vspace{1em} 
\noindent{\bf Long-term goals}
\begin{itemize}
    \item Expanding on the goals above by adding an extra step of physics discovery with ML models to figure out which experiments to perform.
\end{itemize}

\subsection{Trustworthy, robust, interpretable, and explainable AI for quantum technologies}

Key considerations in applying ML to quantum science and technology are trust, robustness, interpretability, and explainability. While \emph{neural networks} 
(NNs) have shown their power in various applications, their lack of transparency hinders the safe and reliable application of these algorithms to valuable quantum systems. Striking a delicate balance between leveraging advanced algorithms and mitigating risks is crucial for instilling confidence in automated control systems. 

Additionally, interpretability and explainability of ML in quantum science are critical for uncovering decision mechanisms used by NNs when addressing complex quantum problems~\cite{wetzel2025interpretable}. To drive scientific discovery, it is vital to not only comprehend the outputs generated by ML algorithms, but to also understand the underlying principles and concepts that guide their reasoning~\cite{krenn2022scientific}. Understanding the factors contributing to a model's predictions allows scientists to assess reliability, validate solutions, and identify biases and errors in training data. Ultimately, this improves the robustness of quantum simulations and predictions. Further, extracting human-understandable knowledge from ML models is pivotal for driving breakthroughs in quantum science and technology. Efficient implementation of approaches that effectively contribute to validating findings, uncovering novel insights, and advancing quantum science through new discoveries is a significant challenge. 


\subsubsection{Interpretable architecture for quantum data analysis}
\label{sec:ae}

One of the most widely used architectures for interpretable ML on quantum data is the \emph{autoencoder} (AE). This model is trained in an unsupervised manner to reconstruct its input, with an informational bottleneck that enforces data compression into a latent space. Variants such as the \emph{variational autoencoder} (VAE) further introduce additional regularization to the latent space, encouraging minimal representations. To successfully reconstruct the input, the AE must learn an efficient encoding which, in the context of physical systems, amounts to extracting the relevant physical features. In quantum science, such architectures have been employed to recover minimal representations of qubit systems from tomographic data~\cite{iten2020discovering,MANDy,nautrup2022operationally, rocchetto2018learning}. Another promising direction involves the use of specifically designed interpretable layers, enabling, e.g., the identification of relevant correlations from snapshots of quantum many-body systems on a lattice~\cite{miles2021correlator}. The latter example points to a broader insight: meaningful analysis of quantum data requires adapting standard ML architectures to its unique properties. This becomes even more crucial in interpretable settings, where extracting the correct interpretation of the input depends directly on the model’s inductive biases and design.

\subsubsection{Interpretable architectures for reinforcement learning in quantum environments}
\label{sec:xrl}

Reinforcement learning is emerging as a powerful tool to solve problems in quantum physics (see Section~\ref{sec:discover_hardware}). While RL algorithms learn to solve increasingly complex problems, interpreting the solutions they provide becomes ever more challenging. 
For example, quantum control tasks (see Section~\ref{sec:automatedcontrol}) may involve sequential decision processes in which a controlling agent must choose certain control actions. These actions may have long-term dependencies, where choices made earlier on have consequences at a (much) later time. These nontrivial, long-term strategies are encoded in the agent’s policy, which is, ultimately, a black box that is not easily understood. 

\emph{Explainable reinforcement learning} (XRL; see, e.g., Refs.~\cite{Puiutta2020,Vouros2022}) is an emerging field of research that aims to explain RL agents and environments. Especially in the quantum domain, where tasks are complex and unintutitive, we can expect that RL strategies will quickly become untractable. However, by integrating XRL methods, researchers can not only solve complex problems but also improve their understanding of the involved processes~\cite{Trenkwalder2023,Melnikov2018}.
\\

\noindent{\bf Short-term goals}
\begin{itemize}
    \item Development of interpretable methods capable of handling the complex properties of quantum data.
    \item Rediscovery of known order parameters, phase spaces, and other relevant physical features directly from experimental data.    
    \item Leverage explainability methods to have models explain choices in quantum environments such as quantum control.
\end{itemize}
\vspace{1em}
\noindent{\bf Mid-term goals}
\begin{itemize}
    \item Discovery of new phases of matter and/or theoretical descriptions of experimental systems by means of interpretable ML-guided research.   
    \item Transfer robustness metrics and methods from safe AI methods to models for quantum data.    
    \item Develop a framework for testing and extracting learned policies of trained agents in a standardized setting.
    \item Integrate interpretable methods to build agents that are both explainable and interpretable.
    \item Leverage insights from safe AI methods into agents that control quantum experiments.
\end{itemize}
\vspace{1em}
\noindent{\bf Long-term goals}
\begin{itemize}
 \item Integration of interpretable architectures with other ML modules, e.g., ML agents controlling an experiment, allowing them to be intrinsically guided toward better interpretations of the data they acquire.
\end{itemize}

\section{Foundational questions}

While much of the current interest in ML and AI stems from applied and practical considerations, fundamental questions about learning systems have in fact been a driving force of much of the progress. The importance of fundamental physics perspectives on traditional ML, which have been critical in achieving the current unprecedented points of accuracy, has been acknowledged even by the 2024 Nobel Prize in physics. Statistical mechanics has led to the development of energy-based techniques and, then, diffusion models have achieved game-changing results in terms of data generation. Insights from studies on the renormalization group and mean-field theory have been used in modeling ML/AI behavior. 

Given the fundamental role that classical physics had in the development of classical AI and ML, it is tempting to ask whether novel QML paradigms can be developed from foundational studies in quantum physics and related disciplines. For instance, quantum statistical mechanics and quantum thermodynamics may have a role as crucial as that of classical statistical mechanics for classical AI. The understanding of correlations in dynamical quantum many-body systems may inspire novel quantum algorithms and approximation methods. Progress on foundational questions can also help defining novel approaches to deal and interact with ``data'' in the quantum world and to better understand learning in a fully quantum setting. Example questions include the different interpretations of quantum physics, the interplay between the quantum and the classical world, the role of the measurement postulate and all the limitations that come from that, such as the no-cloning theorem, but also the emergence of the wave-function collapse from other, more fundamental theories. 

In this section, we speculate on not-fully-understood and yet-to-be discovered connections between physics, (quantum) ML, and AI.\\

\noindent{\bf General goals}
\begin{itemize}
    \item Explore novel QML paradigms inspired by foundational studies.
    \item Develop a fully quantum AI model, leveraging principles like superposition, entanglement, and non-locality to redefine learning in a quantum world.
    \item Better understand learning in the quantum realm, addressing fundamental questions such as the role of no-cloning theorem, quantum measurements, and wave-function collapse.
    \item Investigate quantum learning as a physical process, e.g., whether interacting quantum many-particle systems can naturally perform learning tasks through their natural dynamics.
    \item Apply physics principles to address safety, robustness, explainability, and interpretability in quantum AI systems, ensuring they are reliable and aligned with desired outcomes.
    \item Investigate quantum analogs of classical statistical-mechanical models (e.g., Hopfield networks) and their potential for storing and retrieving quantum information.
    \item Address challenges in the quantum agent-environment paradigm, such as entanglement between agents and environments, and develop frameworks for quantum generalizations of Markov decision processes.
\end{itemize}

\subsection{Physics and (quantum) machine learning}

In this line of investigation, we reflect on the possible new connections between physics and learning systems. The question of new connections can be raised by considering the cutting-edge and future challenges of even classical ML.

\subsubsection{Toward general AI}

Most progress in ML has been achieved by giving up on the dream of general AI, and by focusing on specific sub-tasks.

Recent developments in, e.g., large language models, foreshadow the return to the original challenges, where complicated learning systems (which are not just a large neural network, but a complex transformer architecture) fulfill ever more complicated roles. It is a question whether new ideas in physics modeling of complex systems can lead to new insights.

\subsubsection{Safe AI: robustness, alignment, etc}

While the previous decades were hyper-focused on performance in terms of accuracy, it is abundantly clear that the next phases will require a more complicated metric. Arguably the most important features of AI we are interested in pertain to the safety of AI systems, in various contexts. The simplest cases include verification tasks (ensuring an AI system will never fail in some catastrophic way), robustness (stability to small random or adversarial perturbations of inputs), explainability (the capacity to reason about why certain decisions were made), and so on. It is a question whether physics principles and logic can be used to shed light on the limitations and perspectives of achieving AI systems which are safe.

\subsubsection{Quantum AI}

Quantum ML and AI methods, being developed in the context of quantum information, are already naturally connected to aspects of quantum physics. However, it is unclear whether other parts of quantum theory, such as quantum thermodynamics, quantum statistical mechanics, condensed matter, contains ideas that can be used to elucidate the learning processes of quantum mechanical systems. This remains another challenge.

\subsection{Machine learning and AI in a quantum world}

Another fundamental question stems from the very definition of QML. 
This field does not merely lie at the intersection of quantum information, quantum physics, ML, and AI, but rather somewhere in their union. In this subsection, we  attempt to address a speculative use case where everything can be quantum in nature: the model, the training and inference algorithms, the data, and possibly even the labels. 
Therefore, it is crucial to foster the creation of methodologies and techniques that address quantum AI at a level that is fully encompassed within quantum physics. Ultimately, ML can be understood as information processing. Quantum information can be very different from classical information. Genuine quantum principles and phenomena such as superposition, entanglement, non-locality, contextuality, and other quintessential facets of quantum information may alter both the definition and the meaning of learning in a fully quantum world.

The first class of foundational issues in QML arises when we consider various ML modalities (supervised, unsupervised, reinforcement), and allow data (inputs) and or labels (outputs) to be promoted to genuinely quantum states~\cite{aimeur2006world}. The long-term objective here is to build a fully quantum AI model where all data, training algorithm and inference system are fully quantum.
We identify a number of classes of questions.

\subsubsection{How to define learning}

For simplicity we illustrate the questions on the cases of supervised learning and reinforcement learning. Early investigations into supervised quantum learning date back to the 2000s, where multiple copies of quantum states were considered as inputs~\cite{sasaki2002matching}. This setting has been explored extensively, yielding many interesting results. However, the exact similarities and differences between classical and quantum supervised learning remain unclear. For instance, 
it is still unknown what theoretical or practical limitations might exist when working with 
quantum data sets~\cite{PowerOfData,Maragkopoulos2025} or when mapping classical supervised learning problems onto quantum analogs~\cite{banchi2024statistical}.  If the learner uses up all the quantum data during the training phase, then the learning process is essentially classical, as the training set becomes a classical map.

Hybrid strategies, where classical ML methods are paired with quantum measurements to extract information from quantum data, may be problematic, at least in the worst case, due to exponential complexity of full tomographic methods. Recent advances in less general, but more efficient, methods based on, e.g., classical shadows~\cite{huang2020shadow}, quantum kernel methods, quantum Boltzmann machines or tensor networks, may be efficient for specific tasks, but the general principle is still missing. 

Learning strategies based on quantum memories are more likely to achieve provable speed-ups~\cite{huang2022copies}. Extra care must be taken if the final learned model is stored in a quantum memory, as the latter can neither be copied nor broadcasted. However, there are intriguing cases—such as gentle measurement techniques in shadow tomography—that suggest quantum states can retain usefulness even after partial measurement. Determining scenarios where supervised quantum learning algorithms must retain quantum training data for future tasks remains an open foundational problem. More fundamental problems emerge when we consider genuinely quantum label sets. This scenario introduces significant conceptual challenges.

\subsubsection{Learning as a physical process}

Every quantum algorithm can be expressed as a suitably discretized evolution of a physical process, e.g., with time-dependent Hamiltonians or maps. A natural question is then whether we can define an interacting quantum many-particle system, e.g., with many qubits or other physical particles, whose natural evolution performs the learning task in a fully quantum world. Different ``natural'' evolutions are possible, e.g., via unitary dynamics or via an adiabatic evolution where the system approximately remains in the ground state of some evolving Hamiltonian. A possible example of this research line is about finding quantum analogies to the relationship between classical statistical-mechanical models and associative memories (Hopfield networks). Specifically, it is not clear if quantum channels (Hamiltonians, or open systems) can be similarly used to store and retrieve quantum information. Speculative examples may include storing/retrieving information in ground states of some complex quantum Hamiltonian (e.g., the transverse Ising model, where non-commutative terms enable a rich phase diagram), or in non-local correlations that are created in the scrambling dynamics of chaotic quantum systems. 

\subsubsection{Agent-environment paradigm}

Reinforcement learning, and more generally the learning agent-environment paradigm, also encounters numerous challenges when ``quantized''. In a fully quantum RL setting, both the environment and the agent may become entangled over the course of their interactions. This entanglement complicates the very notion of a ``history'' of interactions because measurements of this history collapse superpositions and potentially interfere destructively with the learning process~\cite{dunjko2015history}. One partial solution is to redefine learning in these contexts using quantum generalizations of \emph{Markov decision processes} (MDPs), as seen in works on quantum observable MDPs~\cite{barry2014QOMDP}.

However, these frameworks typically assume quantum actions are represented as classical descriptions of quantum operations, rather than as quantum states themselves. A more general formalism—allowing agents to generate and act with quantum states—remains underexplored. Quantum processors, e.g., based on port-based teleportation, can be used to write the program in a quantum state, but no efficient training method is known to date~\cite{banchi2020convex}. The question of information extracted during learning is also crucial. If an agent gathers information from a quantum environment, it must operate within the constraints of quantum mechanics—e.g., the no-cloning theorem. This raises fascinating foundational questions: can knowledge itself be defined in terms of quantum states that cannot be shared or copied? While this notion aligns with quantum mechanics, it challenges classical intuitions about knowledge transfer and collaboration in ML.

Ultimate questions here go far, from the limits of quantum autonomous agents to learn new quantum physics, to, in principle, the influence learning may have on foundations of quantum mechanics, e.g., in the definition of an observer which performs a measurement.

\section{Building bridges between quantum and AI}

We believe it is now the right time to invest in research at this emerging interface of quantum science and ML so that the EU can remain competitive with the US, Canada, and China in developing next-generation quantum technology. Patents for ML applications in quantum computing are already picking up speed, but mostly in the US. The quantum flagship has put Europe in a strong position; a broad funding initiative for ML in quantum science will enable Europe to take on the lead in these new developing technologies.

Funding needs to be both for fundamental and applied research projects, in order to cover the full spectrum of developments. While the applications for optimal control are already being prepared for commercial exploitation by the first start-ups, ab initio computational methods are in a more exploratory phase, which requires funding of purely fundamental research for unleashing the full potential. The same holds for other applications such as the analysis of quantum data arising from experiments that might turn out seminal for the understanding of physics or for future technologies.

Facilitating the exchange between the ML and the quantum physics communities has the potential to transform both fields and interdisciplinary teams are needed to push beyond current boundaries. At the core of this proposed initiative is the merging of diverse communities, to bring together a heterogeneous range of views and ensure openness and diversity. For quantum science and technology to synergize with the field of ML and AI, we need to bring together quantum experimentalists, quantum theorists, ML engineers, and computer scientists, but also entrepreneurs and investors.

Open-source software, freely available and standardized benchmark data sets, model databases, and community challenges were central for the rapid advancement of ML techniques. Building on this experience, we believe that creating a similar ecosystem for ML in quantum science will likewise boost progress by removing barriers for interdisciplinary collaboration and optimally tapping the available potential. To this end, it is necessary to standardize quantum physics problems through interoperable and structured interfaces. Their role will be to enable sharing of experimental data and translating quantum physics problems into a common ML language. On the one hand, standardization will enhance the applicability of ML methods in both theoretical and experimental quantum physics, thus improving reusability, reproducibility, and comparability. On the other hand, the development of community-driven projects will create shared spaces, which provide interfaces as tutorials or documentations that help students and researchers to familiarize and strengthen cohesion between fields, and encourage interdisciplinary collaboration and cross fertilization.

Progress in this rapidly developing field requires the training of a next generation of researchers with expertise in quantum science and ML, e.g., via suitable doctoral networks. Additional training and educational resources, such as dedicated online platforms and training resources, and encouraging cross-field conferences and symposia, will simplify the access to state-of-the-art ML and further encourage its widespread adoption by quantum scientists. This can bridge the gap between theory and experiment, by facilitating a more seamless integration of theoretical modeling and experimental data analysis. Such educational programs at the interface of quantum science and ML will produce a workforce that is highly skilled in both forward-looking fields. This is not only fruitful for fundamental research, but also essential to keep replenishing industry with open-minded experts who transfer knowledge into competitive products and services.

Social media and science-communication strategies, as well as collaborations with creators, developers, and industry partners, will play a key role in making ML techniques in quantum physics beneficial to all of society. By placing engagement at the center of the research process, we shall bridge boundaries between disciplines, facilitate the exchange of valuable knowledge with industry partners and policy makers, and improve the public perception of quantum science.

\section{Recommendations and challenges}

\subsection{Theoretical work}

Quantum AI is still a relatively nascent domain. More theoretical insights are needed to better understand how to best employ quantum phenomena to accelerate computing: determining theoretical and practical computing speed-ups, reducing requirements for training data, and obtaining better results. Also, more theoretical work is needed to create efficient bridges between quantum and classical AI, e.g., when a quantum computer is used to train a machine or deep learning model that is then run classically, like in embedded systems (car vision, etc). This work should also cover both fundamental and applied research and practical use cases, e.g., in healthcare.

\subsection{Aligning with quantum hardware road maps}

Implementing quantum-assisted AI algorithms is highly dependent on the progress of quantum computing hardware. Particularly, it is related to the quality and quantity of available qubits. 
Quantum AI capabilities will progress synchronously with hardware evolutions, such as advanced NISQ devices with better qubit fidelities \cite{bharti_2021_noisy}, \emph{early fault-tolerant quantum computers} (eFTQC) with about a hundred logical qubits, and beyond, with utility-scale FTQCs supporting thousands of logical qubits. Algorithmic advances in quantum AI will guide quantum hardware academics and industry vendors in adjusting their road maps. Likewise, quantum ML developers will synchronize their work with hardware vendors.
A second aspect deals with qRAM (\emph{quantum random access memory}) research \cite{PhysRevLett.100.160501}. qRAM might arguably be an important enabling technology for quantum machine learning, particularly to address the pressing challenges with data preparation and loading.

\subsection{Estimating resources}

In relation to the previous point, research work on quantum AI as well as on the usage of classical ML for the development of quantum computing hardware and software must rely on careful resource analysis. Such analysis will help identify scaling issues and avoid massive energetic needs that society will struggle to meet. This transversal research will benefit from the involvement of the EuroHPC participating organizations.

As concern with the energy consumption of current AI and LLM solutions is growing, some academic and industry work should estimate, benchmark and optimize the energy consumption of both quantum AI solutions and classical AI tools used in quantum technologies.

When AI is used as a tool for calibration, error analysis for mitigation or correction, or other enabling tasks in quantum computing, it is crucial that its energy consumption does not negate the energy savings expected in quantum computing relative to classical HPC. 

\subsection{Engaging classical AI specialists}

Advancing the field of quantum machine learning as well as the use of ML in the context of quantum computing requires more engagement of the classical AI scientific community. Cross-discipline initiatives may be launched in education and community buildouts to encourage it.
This will also enable the quantum community to better qualify the challenges ahead with classical AI.

\subsection{Software engineering}

The development of EU-based know-how and competitive advantages in classical and quantum AI should lead to the development of new software engineering tools. Such tools include quantum code compilers and optimizers leveraging ML, tools enabling quantum code debugging, and the like. Proper usage of LLM-based software engineering will also be crucial for increasing the productivity of quantum software developers.

\subsection{Open science and industry competitiveness}

The EU is highly challenged by US dominance in the AI field, at the hardware level (Nvidia) as well as with software and cloud infrastructures (OpenAI, Google, Meta, AWS). Meanwhile, a vibrant innovation ecosystem works well with open science and research processes. We have in mind that open science is needed to advance the field while developing the quantum industry ecosystem.

Managing this delicate equilibrium requires robust EU and member states fundamental research funding, requiring that resulting publications and data sets are openly accessible. This ensures a steady flow of new discoveries leveraged for the development of commercial ventures. It goes with an effective \emph{intellectual property} (IP) framework that allows researchers to publish freely while securing IP and the launch of spin-offs startups. Moreover, adopting an open innovation mindset, in which pre-competitive platforms and standards are developed collaboratively, enables companies to compete on final products, services and business models, while benefiting from shared R\&D. Academic–industry partnerships, joint research centers, and private-sector sponsored fellowships can contribute to enriching the talent pool and nurture knowledge transfer without locking valuable insights behind corporate walls. Working on standardization and benchmarking tools can also contribute to shape the competitive landscape.

EU funding instruments can contribute to this synergy with individual (ERC) and collaborative research grants (EU Quantum Flagship), proof-of-concept grants (EIC), dedicated incubators, and large-cap funding (EIB).

\subsection{Education}

The computer science education landscape is currently dominated by AI. In order to enable the future programs related to this white paper, EU member states will need to train more scientists at the crossroads of AI and quantum computing. New curricula should be proposed to develop the skills of quantum-AI scientists and engineers.

\subsection{Societal challenges}

Quantum technologies and AI are strategic technology fields for the EU. They are now widely addressed at dedicated industry fairs and investor meetings. While quantum technologies themselves can be considered as neutral tools, the societal challenges largely arise from how these technologies are applied. The EU is already tackling the societal impact of these emerging technologies. Indeed, the \emph{General Data Protection Regulation} (GDPR) and the AI Act regulate many aspects related to data protection, transparency, and accountability in high-tech fields, including QT applications. 

Furthermore, existing initiatives in Europe are addressing the societal challenges of QT such as the Quantum Delta Centre for quantum and society in the Netherlands, the QuantWorld project in Germany, the Innsbruck Quantum Ethics Lab in Austria, and the Humanities for Quantum Sciences lab in France.

As joint efforts work best when parties coming from different disciplines can develop a common language, the expertise of \emph{societal sciences and humanities} (SSH) may play a significant role in the joint development of AI and quantum technologies. It will ensure that societal and ethical considerations are embedded within the broader strategic development of quantum technologies in Europe.

\section*{Acknowledgments}

The contributors would like to thank Artur Garcia, Laure Le Bars, Philippe Grangier, Thierry Debuisschert, Phila Rembold, Giuseppe Carleo, Marcus Huber, Annabelle Bohrdt and Sebastian Erne.
\\

This manuscript's preparation was coordinated by Enrique Sánchez-Bautista (Quantum Community Network---Brussels Office), Monica Constantin (Quantum Community Network---Brussels Office), and Ziyad Amodjee (CNRS).

\hfill 

The bibliographical references from this work have been selected by its coauthors based on their relevance and citations with a mix of foundational and review papers. Use cases references are either preprints or peer-reviewed manuscripts.

\section*{List of acronyms}
\begin{description}[noitemsep]
\item[AE] Autoencoder
\item[AI] Artificial intelligence
\item[CFN] Collision-free navigation
\item[DFT] Density functional theory
\item[EIB] European Innovation Bank
\item[EIC] European Innovation Council
\item[ERC] European Research Council
\item[eFTQC] Early fault-tolerant quantum computer
\item[FTQC] Fault-tolerant quantum computer
\item[GDPR] General data protection regulation
\item[HHL] Harrow--Hassidim--Lloyd
\item[HEP] High-energy physics
\item[HPC] High-performance computing
\item[IP] Intellectual property
\item[LLM] Large language model
\item[MAS] Multi-agent system
\item[MDP] Markov decision process
\item[MILP] Mixed-integer linear programming
\item[ML] Machine learning
\item[MRI] Magnetic resonance imaging
\item[NISQ] Noisy intermediate-scale quantum
\item[NN] Neural network
\item[NQS] Neural quantum states
\item[PCA] Principal component analysis
\item[POMDP] Partially observable Markov decision process
\item[PQC] Parameterized quantum circuit
\item[QAI] Quantum artificial intelligence
\item[QAOA] Quantum approximate optimization algorithm
\item[QCNN] Quantum convolutional neural network
\item[QCV] Quantum computer vision
\item[QEC] Quantum error correction
\item[QEM] Quantum error mitigation
\item[QFT] Quantum Fourier transform
\item[QHD] Quantum Hamiltonian descent
\item[QMAS] Quantum multi-agent system
\item[QMDP] Quantum Markov decision process
\item[QML] Quantum machine learning
\item[QNLP] Quantum natural language processing
\item[QNN] Quantum neural network
\item[QP] Quantum-supported planning
\item[QPS] Quantum-supported/automated planning and scheduling
\item[QR] Quantum reasoning
\item[QS] Quantum-supported scheduling
\item[qRAM] Quantum random access memory
\item[QRL] Quantum reinforcement learning
\item[QSL] Quantum supervised learning
\item[QT] Quantum technology
\item[QUBO] Quadratic unconstrained binary optimization
\item[QV] Quantum vision computing
\item[R\&D] Research and development
\item[RL] Reinforcement learning
\item[SRIA] Strategic Research and Industry Agenda
\item[SSH] Social sciences and humanities
\item[VAE] Variational autoencoder
\item[VQA] Variational quantum algorithm
\item[XRL] Explainable reinforcement learning
\end{description}

\bibliography{biblio}

\end{document}